\documentclass[11pt]{article}
\usepackage{graphicx,amsmath,amssymb,%citesort,
cite,epsfig,epsf}
\usepackage{fullpage}
\usepackage[small,bf]{caption}
\def\real    { \mathbb{R} }
\def\gaussian { \mathcal{N} }
\newcommand{\norm}[1]{\left\| #1 \right\|}
\def\Pzero{\mathrm{P_0}}
\def\Pone{\mathrm{P_1}}
\def\WPzero{\mathrm{WP_0}}
\def\WPone{\mathrm{WP_1}}

\def\xhat{x}
\def\sol{x^\star}
\def\true{x_0}

\def\alphahat{\alpha}

\newcommand{\ncol}{n}
\newcommand{\nrow}{m}
\newcommand{\nsparse}{k}
\newcommand{\R}{\mathbb{R}}

\title{Enhancing Sparsity by Reweighted $\ell_1$ Minimization}

\author{Emmanuel J. Cand\`es$^{\dagger}$
\and Michael B. Wakin$^{\sharp}$
\and Stephen P. Boyd$^{\S}$\\
  \vspace{-.1cm}\\
  $\dagger$ Applied and Computational Mathematics,
  Caltech, Pasadena, CA 91125\\
  \vspace{-.3cm}\\
  $\sharp$ Electrical Engineering \& Computer Science, University
  of Michigan, Ann Arbor, MI, 48109\\
  \vspace{-.3cm}\\
  $\S$ Department of Electrical Engineering, Stanford University,
  Stanford, CA 94305}

\date{October 2007}

\begin{document}

\maketitle

\begin{abstract}
  It is now well understood that (1) it is possible to reconstruct
  sparse signals exactly from what appear to be highly incomplete sets
  of linear measurements and (2) that this can be done by constrained
  $\ell_1$ minimization.  In this paper, we study a novel method for
  sparse signal recovery that in many situations outperforms $\ell_1$
  minimization in the sense that substantially fewer measurements are
  needed for exact recovery. The algorithm consists of solving a
  sequence of weighted $\ell_1$-minimization problems where the
  weights used for the next iteration are computed from the value of
  the current solution.  We present a series of experiments
  demonstrating the remarkable performance and broad applicability of
  this algorithm in the areas of sparse signal recovery, statistical
  estimation, error correction and image processing. Interestingly,
  superior gains are also achieved when our method is applied to
  recover signals with assumed near-sparsity in overcomplete
  representations---not by reweighting the $\ell_1$ norm of the
  coefficient sequence as is common, but by reweighting the $\ell_1$
  norm of the transformed object. An immediate consequence is the
  possibility of highly efficient data acquisition protocols by
  improving on a technique known as compressed sensing.
\end{abstract}


\section{Introduction}
\label{sec:intro}

What makes some scientific or engineering problems at once interesting
and challenging is that often, one has fewer equations than
unknowns. When the equations are linear, one would like to determine
an object $\true \in \R^\ncol$ from data $y = \Phi \true$, where
$\Phi$ is an $\nrow \times \ncol$ matrix with fewer rows than columns;
i.e., $\nrow < \ncol$. The problem is of course that a system with fewer
equations than unknowns usually has infinitely many solutions and
thus, it is apparently impossible to identify which of these candidate
solutions is indeed the ``correct'' one without some additional
information.

In many instances, however, the object we wish to recover is known
to be structured in the sense that it is sparse or compressible.
This means that the unknown object depends upon a smaller number of
unknown parameters. In a biological experiment, one could measure
changes of expression in 30,000 genes and expect at most a couple
hundred genes with a different expression level. In signal
processing, one could sample or sense signals which are known to be
sparse (or approximately so) when expressed in the correct basis.
This premise radically changes the problem, making the search for
solutions feasible since the simplest solution now tends to be the
right one.

Mathematically speaking and under sparsity assumptions, one would
want to recover a signal $\true \in \R^\ncol$, e.g.,~the coefficient
sequence of the signal in the appropriate basis, by solving the
combinatorial optimization problem
\begin{equation}
  (\Pzero) \qquad \min_{\xhat \in \R^{\ncol}} \,
\|\xhat\|_{\ell_0} \quad \textrm{subject to} \quad  y = \Phi \xhat,
\label{eq:ell0opt}
\end{equation}
where $\|x\|_{\ell_0} = |\{i : x_i \neq 0\}|$. This is a common sense
approach which simply seeks the simplest explanation fitting the data.
In fact, this method can recover sparse solutions even in situations in which
$\nrow \ll \ncol$. Suppose for example that all sets of $\nrow$
columns of $\Phi$ are in general position.  Then the program
($\Pzero)$ perfectly recovers all sparse signals $\true$ obeying
$\|\true\|_{\ell_0} \le \nrow/2$. This is of little practical use,
however, since the optimization problem \eqref{eq:ell0opt} is
nonconvex and generally impossible to solve as its solution usually
requires an intractable combinatorial search.

A common alternative is to consider the convex problem
\begin{equation}
  (\Pone) \qquad \min_{\xhat \in \R^{\ncol}} \,
\|\xhat\|_{\ell_1} \quad \textrm{subject to} \quad  y = \Phi \xhat,
\label{eq:ell1opt}
\end{equation}
where $\|\xhat\|_{\ell_1} = \sum_{i = 1}^\ncol |x_i|$.  Unlike
($\Pzero$), this problem is convex---it can actually be recast as a
linear program---and is solved efficiently \cite{BoV:04}.
The programs ($\Pzero$)
and ($\Pone$) differ only in the choice of objective function, with
the latter using an $\ell_1$ norm as a proxy for the literal $\ell_0$
sparsity count. As summarized below, a recent body of work has shown
that perhaps surprisingly, there are conditions guaranteeing a formal
equivalence between the combinatorial problem ($\Pzero$) and its
relaxation $(\Pone)$.

The use of the $\ell_1$ norm as a sparsity-promoting functional
traces back several decades. A leading early application was
reflection seismology, in which a sparse reflection function
(indicating meaningful changes between subsurface layers) was sought
from bandlimited data. In 1973, Claerbout and
Muir~\cite{claerbout73ro} first proposed the use of $\ell_1$ to
deconvolve seismic traces. Over the next decade this idea was
refined to better handle observation noise~\cite{taylor79de,
santosa86li}, and the sparsity-promoting nature of $\ell_1$
minimization was empirically confirmed. Rigorous results began to
appear in the late-1980's, with Donoho and Stark~\cite{donoho89un}
and Donoho and Logan~\cite{donoho92si} quantifying the ability to
recover sparse reflectivity functions. The application areas for
$\ell_1$ minimization began to broaden in the mid-1990's, as the
LASSO algorithm~\cite{tibshirani96re} was proposed as a method in
statistics for sparse model selection, Basis Pursuit~\cite{DonohoBP}
was proposed in computational harmonic analysis for extracting a
sparse signal representation from highly overcomplete dictionaries,
and a related technique known as total variation minimization was
proposed in image processing~\cite{RudinOsherFatemi92,colorTV}.

Some examples of $\ell_1$ type methods for sparse design in
engineering include Vandenberghe et al.~\cite{VBE:97,VBE:98} for
designing sparse interconnect wiring, and Hassibi et
al.~\cite{HHB:99} for designing sparse control system feedback
gains. In~\cite{DaD:95}, Dahleh and Diaz-Bobillo solve controller
synthesis problems with an $\ell_1$ criterion, and observe that the
optimal closed-loop responses are sparse. Lobo et al.\ used $\ell_1$
techniques to find sparse trades in portfolio optimization with
fixed transaction costs in~\cite{LFB:06}. In~\cite{GhB:06}, Ghosh
and Boyd used $\ell_1$ methods to design well connected sparse
graphs; in~\cite{SBXD:06}, Sun et al.\ observe that optimizing the
rates of a Markov process on a graph leads to sparsity. In~\cite[\S
6.5.4, \S 11.4.1]{BoV:04}, Boyd and Vandenberghe describe several
problems involving $\ell_1$ methods for sparse solutions, including
finding small subsets of mutually infeasible inequalities, and
points that violate few constraints. In a recent paper, Koh et al.\
used these ideas to carry out piecewise-linear trend
analysis~\cite{KKBG:07}.

Over the last decade, the applications and understanding of $\ell_1$
minimization have continued to increase dramatically. Donoho and
Huo~\cite{donoho01un} provided a more rigorous analysis of Basis
Pursuit, and this work was extended and refined in subsequent years,
see \cite{EladBruckstein,GribonvalNielsen,TroppRelax}. Much of the
recent focus on $\ell_1$ minimization, however, has come in the
emerging field of Compressive
Sensing~\cite{CandesRUP,CandesUES,DonohoCS}. This is a setting where
one wishes to recover a signal $\true$ from a small number of
compressive measurements $y = \Phi \true$. It has been shown that
$\ell_1$ minimization allows recovery of sparse signals from
remarkably few measurements~\cite{tanner06,CandesSSR}: supposing
$\Phi$ is chosen randomly from a suitable distribution, then with very
high probability, all sparse signals $\true$ for which
$\|\true\|_{\ell_0} \le \nrow/\alpha$ with $\alpha =
O(\log(\ncol/\nrow))$ can be {\em perfectly} recovered by using
$(\Pone)$. Moreover, it has been established \cite{CandesSSR} that
Compressive Sensing is robust in the sense that $\ell_1$ minimization
can deal very effectively (a) with only approximately sparse signals
and (b) with measurement noise.  The implications of these facts are
quite far-reaching, with potential applications in data
compression~\cite{CandesUES,DonohoECS}, digital
photography~\cite{CScam06}, medical
imaging~\cite{CandesRUP,lustig07sp}, error
correction~\cite{CandesDLP,ErrorCorrect}, analog-to-digital
conversion~\cite{baa}, sensor networks~\cite{bajwa06co,dcsJournal},
and so on. (We will touch on some more concrete examples in
Section~\ref{sec:exp}.)

The use of $\ell_1$ regularization has become so widespread that it
could arguably be considered the ``modern least squares''.  This
raises the question of whether we can improve upon $\ell_1$
minimization? It is natural to ask, for example, whether a different
(but perhaps again convex) alternative to $\ell_0$ minimization might
also find the correct solution, but with a lower measurement
requirement than $\ell_1$ minimization.

In this paper, we consider one such alternative, which aims to help
rectify a key difference between the $\ell_1$ and $\ell_0$ norms,
namely, the dependence on magnitude: larger coefficients are penalized
more heavily in the $\ell_1$ norm than smaller coefficients, unlike
the more democratic penalization of the $\ell_0$ norm. To address this
imbalance, we propose a weighted formulation of $\ell_1$ minimization
designed to more democratically penalize nonzero coefficients. In
Section~\ref{sec:overview}, we discuss an iterative algorithm for
constructing the appropriate weights, in which each iteration of the
algorithm solves a convex optimization problem, whereas the overall
algorithm does not.  Instead, this iterative algorithm attempts to
find a local minimum of a concave penalty function that more closely
resembles the $\ell_0$ norm. Finally, we would like to draw attention
to the fact that each iteration of this algorithm simply requires
solving one $\ell_1$ minimization problem, and so the method can be
implemented readily using existing software.

In Section~\ref{sec:exp}, we present a series of experiments
demonstrating the superior performance and broad applicability of this
algorithm, not only for recovery of sparse signals, but also
pertaining to compressible signals, noisy measurements, error
correction, and image processing. This section doubles as a brief tour
of the applications of Compressive Sensing. In
Section~\ref{sec:l1analysis}, we demonstrate the promise of this
method for efficient data acquisition.  Finally, we conclude in
Section~\ref{sec:discussion} with a final discussion of related work
and future directions.

\section{An iterative algorithm for reweighted $\ell_1$ minimization}
\label{sec:overview}

\subsection{Weighted $\ell_1$ minimization}

Consider the ``weighted'' $\ell_1$ minimization problem
\begin{equation}
(\WPone) \qquad \min_{\xhat \in \real^\ncol}  \, \sum_{i = 1} w_i |x_i| \quad
\textrm{subject to}  \quad y = \Phi \xhat \label{eq:well1opt},
\end{equation}
where $w_1, w_2, \dots, w_\ncol$ are positive weights.  Just like its
``unweighted'' counterpart ($\Pone$), this convex problem can be
recast as a linear program. In the sequel, it will be convenient to
denote the objective functional by $\|W x\|_{\ell_1}$ where $W$ is the
diagonal matrix with $w_1, \dots, w_\ncol$ on the diagonal and zeros
elsewhere.

The weighted $\ell_1$ minimization ($\WPone$) can be viewed as a
relaxation of a weighted $\ell_0$ minimization problem
\begin{equation}
(\WPzero) \qquad  \min_{\xhat \in \real^\ncol}  \, \|W x\|_{\ell_0} \quad
\textrm{subject to}  \quad y = \Phi \xhat \label{eq:well0opt}.
\end{equation}
Whenever the solution to ($\Pzero$) is unique, it is also the unique
solution to ($\WPzero$) provided that the weights do not vanish.
However, the corresponding $\ell_1$ relaxations ($\Pone$) and
($\WPone$) will have different solutions in general. Hence, one may
think of the weights $(w_i)$ as free parameters in the convex
relaxation, whose values---if set wisely---could improve the signal
reconstruction.

This raises the immediate question: what values for the weights will
improve signal reconstruction? One possible use for the weights could
be to counteract the influence of the signal magnitude on the $\ell_1$
penalty function. Suppose, for example, that the weights were
inversely proportional to the true signal magnitude, i.e., that
\begin{equation}
w_i = \left\{ \begin{array}{cc} \frac{1}{|x_{0,i}|}, & ~~~ x_{0,i} \neq 0, \\
\infty, & ~~~ x_{0,i} = 0. \end{array} \right. \label{eq:optweights}
\end{equation}
If the true signal $\true$ is $k$-sparse, i.e.,~obeys
$\|\true\|_{\ell_0} \le k$, then ($\WPone$) is guaranteed to find
the correct solution with this choice of weights, assuming only that
$\nrow \ge \nsparse$ and that just as before, the columns of $\Phi$
are in general position.  The large (actually infinite) entries in
$w_i$ force the solution $\xhat$ to concentrate on the indices where
$w_i$ is small (actually finite), and by construction these
correspond precisely to the indices where $\true$ is nonzero. It is
of course impossible to construct the precise weights
\eqref{eq:optweights} without knowing the signal $\true$ itself, but
this suggests more generally that large weights could be used to
discourage nonzero entries in the recovered signal, while small
weights could be used to encourage nonzero entries.

\begin{figure}
\begin{center}
\begin{tabular}{ccc}
~~
\includegraphics[scale=0.8]{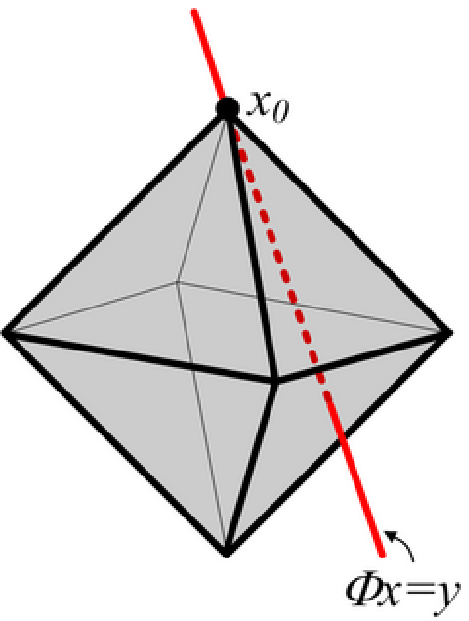}~~
& ~~
\includegraphics[scale=0.8]{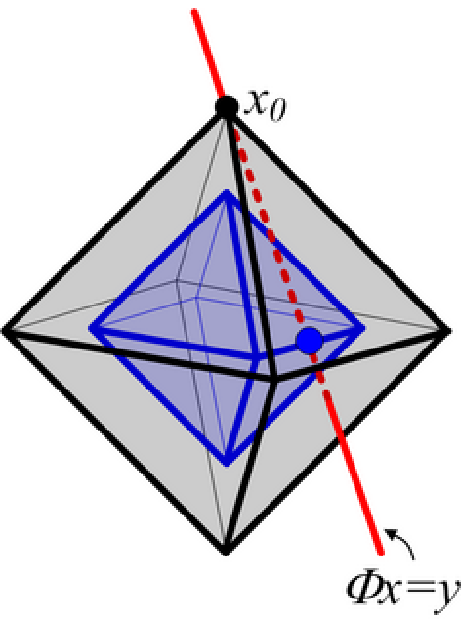}~~  &
~~  \includegraphics[scale=0.8]{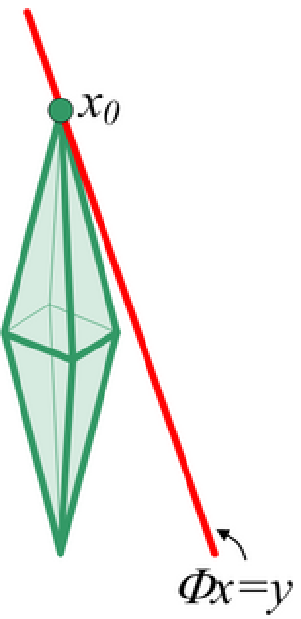}~~\\
(a) &  (b) &  (c)
\end{tabular}
\end{center}
\caption{\small\sl Weighting $\ell_1$ minimization to
improve sparse signal recovery. (a) Sparse signal $\true$, feasible set
$\Phi \xhat = y$, and $\ell_1$ ball of radius $\|\true\|_{\ell_1}$. (b) There
exists an $\xhat \neq \true$ for which $\|\xhat\|_{\ell_1} < \|\true\|_{\ell_1}$. (c)
Weighted $\ell_1$ ball. There exists no $\xhat \neq \true$ for which
$\|W\xhat\|_{\ell_1} \le \|W\true\|_{\ell_1}$.}\label{fig:pinched}
\end{figure}

For the sake of illustration, consider the simple 3-D example in
Figure~\ref{fig:pinched}, where $\true = [0 ~ 1 ~ 0]^T$ and
\[
\Phi = \left[
\begin{array}{ccc} 2 & 1 & 1 \\ 1 & 1 & 2 \end{array} \right].
\]
We wish to recover $\true$ from $y = \Phi \true = [1 ~ 1]^T$.
Figure~\ref{fig:pinched}(a) shows the original signal $\true$, the
set of points $\xhat \in \real^3$ obeying $\Phi\xhat = \Phi \true =
y$, and the $\ell_1$ ball of radius 1 centered at the origin. The
interior of the $\ell_1$ ball intersects the feasible set $\Phi\xhat
= y$, and thus ($\Pone$) finds an incorrect solution, namely,
$\sol = [1/3 ~ 0 ~ 1/3]^T \neq \true$ (see
Figure~\ref{fig:pinched}(b)).

Consider now a hypothetical weighting matrix $W = \mathrm{diag}([3 ~ 1
~ 3]^T)$. Figure~\ref{fig:pinched}(c) shows the ``weighted $\ell_1$
ball'' of radius $\|Wx\|_{\ell_1} = 1$ centered at the
origin. Compared to the unweighted $\ell_1$ ball
(Figure~\ref{fig:pinched}(a)), this ball has been sharply pinched at
$\true$. As a result, the interior of the weighted $\ell_1$ ball does
not intersect the feasible set, and consequently, ($\WPone$) will
find the correct solution $\sol = \true$. Indeed, it is not
difficult to show that the same statements would hold true for any
positive weighting matrix for which $w_2 < (w_1+w_3)/3$. Hence there
is a range of valid weights for which ($\WPone$) will find the
correct solution. As a rough rule of thumb, the weights should relate
inversely to the true signal magnitudes.

\subsection{An iterative algorithm}
\label{sec:alg}

The question remains of how a valid set of weights may be obtained
without first knowing $\true$. As Figure~\ref{fig:pinched} shows,
there may exist a range of favorable weighting matrices $W$ for each
fixed $\true$, which suggests the possibility of constructing a
favorable set of weights based solely on an approximation $\xhat$ to
$\true$ or on other side information about the vector magnitudes.

We propose a simple iterative algorithm that alternates between
estimating $\true$ and redefining the weights. The algorithm is as
follows:
\begin{enumerate}
\item Set the iteration count $\ell$ to zero and $w^{(0)}_i = 1$, $i =
  1, \ldots, \ncol$.
\item Solve the weighted $\ell_1$ minimization problem
\[
\xhat^{(\ell)} = \arg\min \|W^{(\ell)} \xhat\|_{\ell_1}
\quad \textrm{subject to} \quad y = \Phi \xhat.
\]
\item Update the weights: for each $i = 1, \ldots, \ncol$,
\begin{equation}
  w^{(\ell+1)}_i = \frac{1}{|\xhat^{(\ell)}_i|+\epsilon}.
\label{eq:rwrule}
\end{equation}
\item Terminate on convergence or when $\ell$ attains a specified
  maximum number of iterations $\ell_{\mathrm{max}}$. Otherwise,
  increment $\ell$ and go to step 2.
\end{enumerate}
We introduce the parameter $\epsilon > 0$ in step 3 in order to
provide stability and to ensure that a zero-valued component in
$\xhat^{(\ell)}$ does not strictly prohibit a nonzero estimate at the
next step. As empirically demonstrated in Section~\ref{sec:exp},
$\epsilon$ should be set slightly smaller than the expected nonzero
magnitudes of $\true$. In general, the recovery process tends to be
reasonably robust to the choice of $\epsilon$.

Using an iterative algorithm to construct the weights $(w_i)$ tends to
allow for successively better estimation of the nonzero coefficient
locations. Even though the early iterations may find inaccurate
signal estimates, the largest signal coefficients are most likely to
be identified as nonzero. Once these locations are identified, their
influence is downweighted in order to allow more sensitivity for
identifying the remaining small but nonzero signal coefficients.

\begin{figure}
\begin{center}
\begin{tabular}{cc}
\includegraphics[scale=0.4]{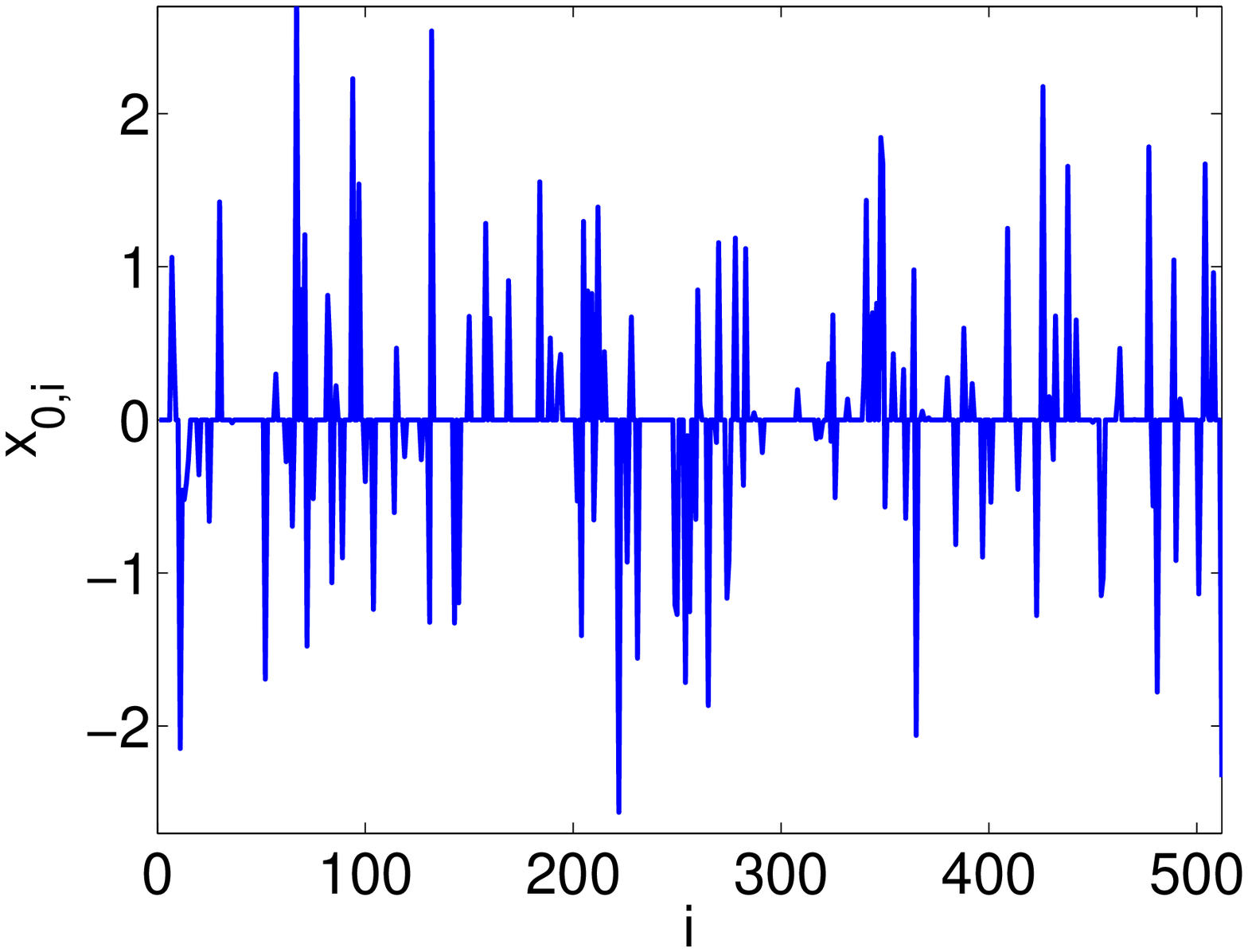} &
\includegraphics[scale=0.4]{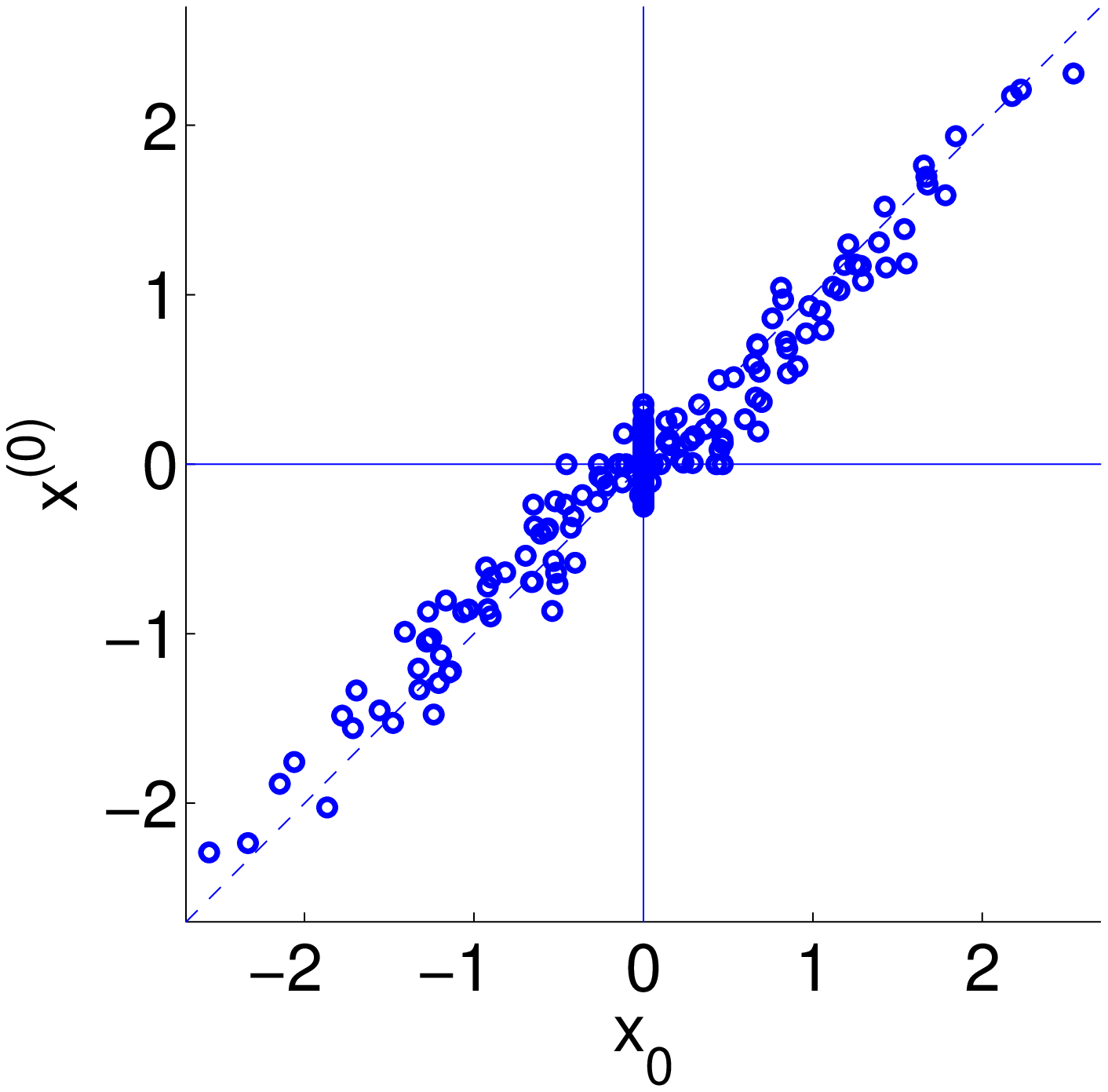} \\
(a) & (b) \\
\includegraphics[scale=0.4]{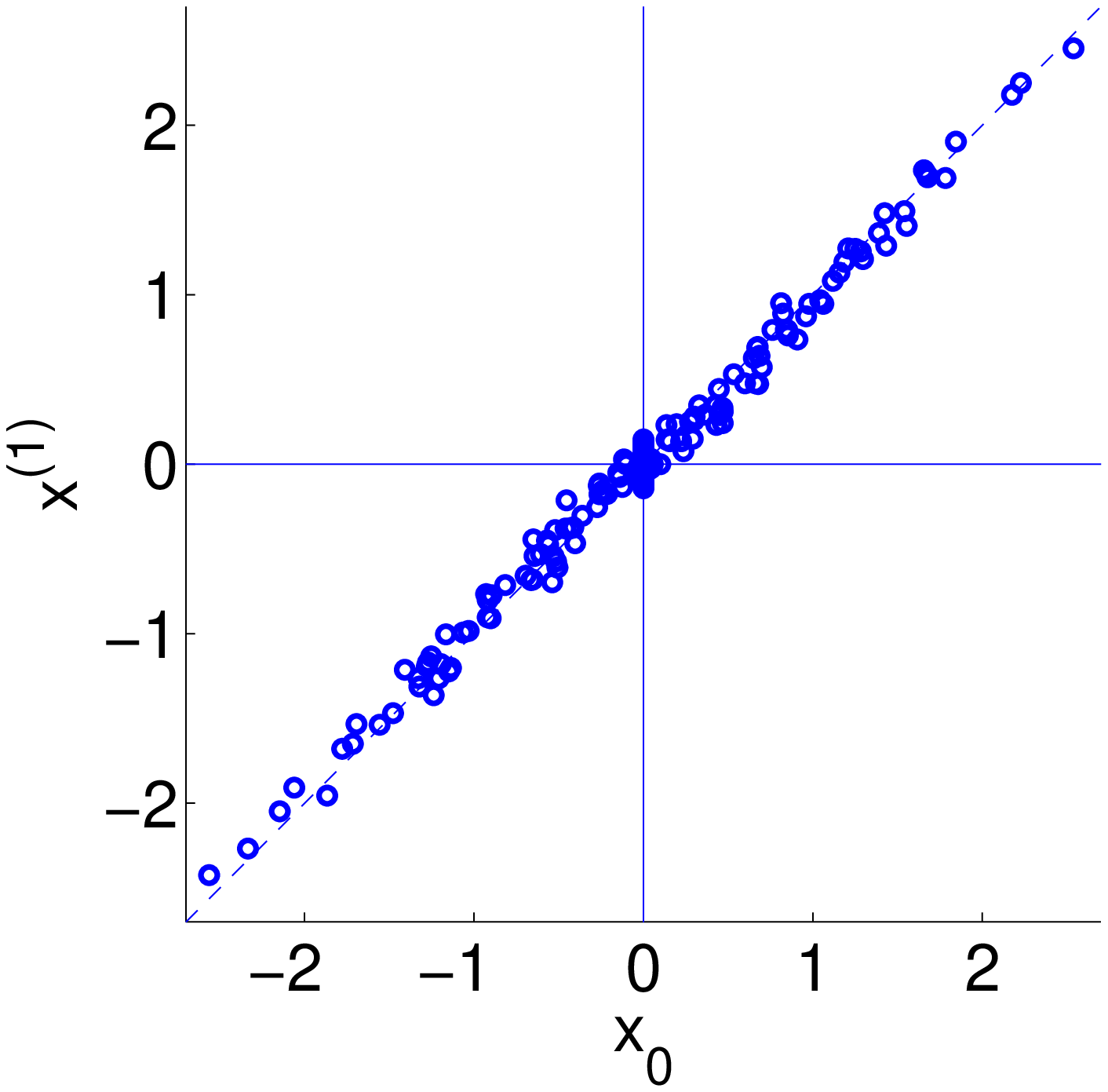} &
\includegraphics[scale=0.4]{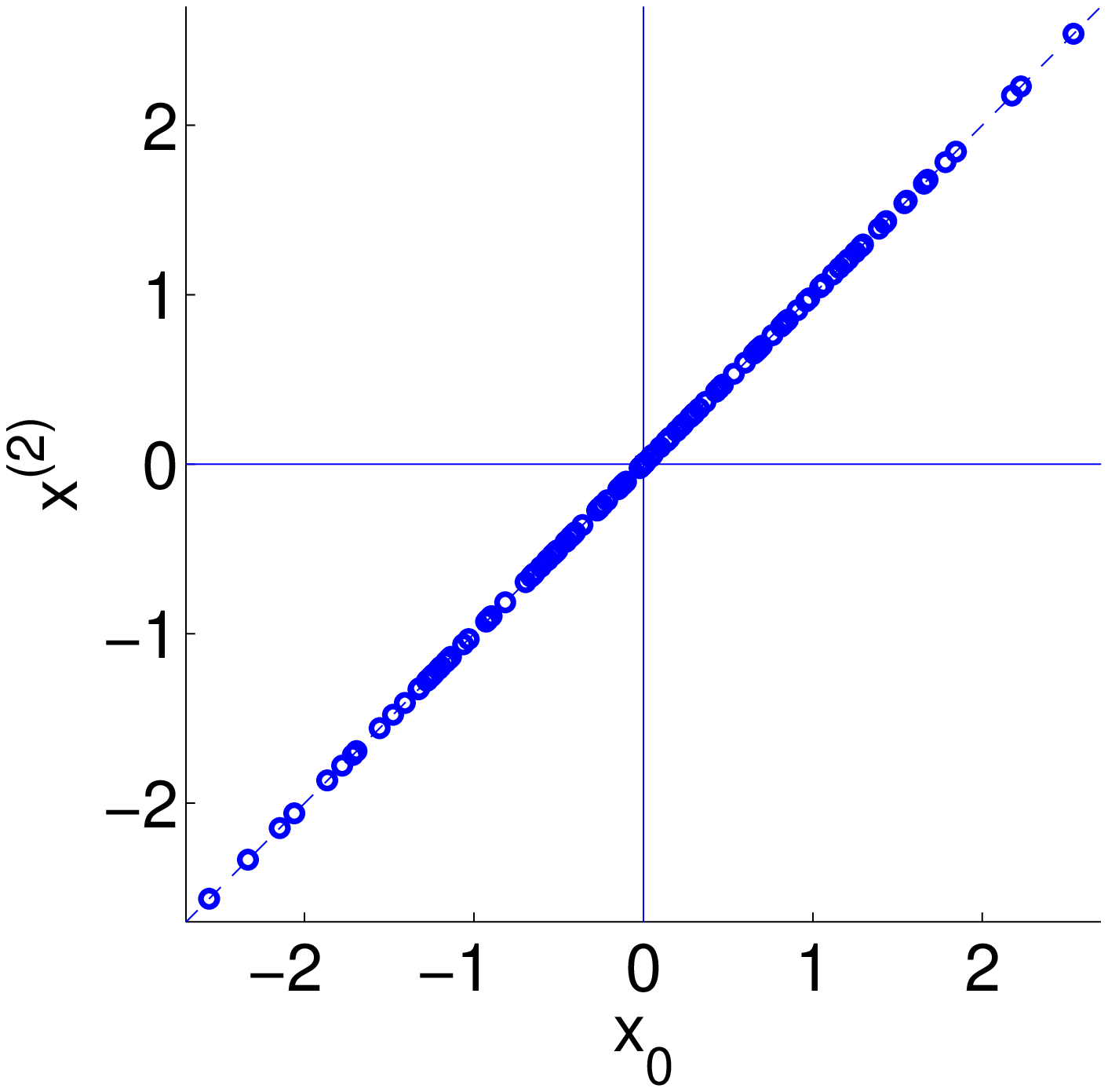} \\
(c) & (d)
\end{tabular}
\end{center}
\caption{\small\sl Sparse signal recovery through reweighted $\ell_1$
  iterations. (a) Original length $\ncol = 512$ signal $\true$ with
  130 spikes. (b) Scatter plot, coefficient-by-coefficient, of $\true$
  versus its reconstruction $\xhat^{(0)}$ using unweighted $\ell_1$
  minimization. (c) Reconstruction $\xhat^{(1)}$ after the first
  reweighted iteration. (d) Reconstruction $\xhat^{(2)}$ after the
  second reweighted iteration.  }\label{fig:est1}
\end{figure}

Figure~\ref{fig:est1} illustrates this dynamic by means of an
example in sparse signal recovery. Figure~\ref{fig:est1}(a) shows
the original signal of length $\ncol=512$, which contains $130$
nonzero spikes. We collect $\nrow = 256$ measurements where the
matrix $\Phi$ has independent standard normal entries. We set
$\epsilon = 0.1$ and $\ell_{\mathrm{max}} = 2$.
Figures~\ref{fig:est1}(b)-(d) show scatter plots,
coefficient-by-coefficient, of the original signal coefficient
$\true$ versus its reconstruction $\xhat^{(\ell)}$. In the
unweighted iteration (Figure~\ref{fig:est1}(b)), we see that all
large coefficients in $\true$ are properly identified as nonzero
(with the correct sign), and that
$\|\true-\xhat^{(0)}\|_{\ell_\infty} = 0.4857$. In this first
iteration, $\|\xhat^{(0)}\|_{\ell_0} = 256 = \nrow$, with 15 nonzero
spikes in $\true$ reconstructed as zeros and 141 zeros in $\true$
reconstructed as nonzeros. These numbers improve after one
reweighted iteration (Figure~\ref{fig:est1}(c)) with now
$\|x-\xhat^{(1)}\|_{\ell_\infty} = 0.2407$,
$\|\xhat^{(1)}\|_{\ell_0} = 256 = \nrow$, 6 nonzero spikes in
$\true$ reconstructed as zeros and 132 zeros in $\true$
reconstructed as nonzeros. This improved signal estimate is then
sufficient to allow perfect recovery in the second reweighted
iteration (Figure~\ref{fig:est1}(d)).

\subsection{Analytical justification}

The iterative reweighted algorithm falls in the general class of MM
algorithms, see \cite{Lange} and references therein. In a nutshell, MM
algorithms are more general than EM algorithms, and work by
iteratively minimizing a simple surrogate function majorizing a given
objective function. To establish this connection, consider the problem
\begin{equation}
  \min_{x \in \real^\ncol} ~ \sum_{i = 1}^\ncol \log(|\xhat_i| + \epsilon)
  \quad \textrm{subject to} \quad y = \Phi \xhat,
\label{eq:logsum}
\end{equation}
which is equivalent to
\begin{equation}
  \min_{x, u \in \real^\ncol} ~ \sum_{i = 1}^\ncol \log(u_i + \epsilon)
  \quad \textrm{subject to} \quad \begin{array}{l} y  = \Phi \xhat, \\
    |\xhat_i| \le u_i, \,\,\, i = 1, \ldots, \ncol.
\end{array}
\label{eq:logsum2}
\end{equation}
The equivalence means that if $x^\star$ is a solution to
\eqref{eq:logsum}, then $(x^\star, |x^\star|)$ is a solution to
\eqref{eq:logsum2}. And conversely, if $(x^\star, u^\star)$ is a
solution to \eqref{eq:logsum2}, then $x^\star$ is a solution to
\eqref{eq:logsum}.

Problem \eqref{eq:logsum2} is of the general form
\begin{equation*}
\label{eq:abstract}
 \min_{v} ~ g(v)   \quad \textrm{subject to} \quad v \in {\cal C},
\end{equation*}
where ${\cal C}$ is a convex set. In \eqref{eq:logsum2}, the function
$g$ is concave and, therefore, below its tangent. Thus, one can
improve on a guess $v$ at the solution by minimizing a linearization
of $g$ around $v$. This simple observation yields the following MM
algorithm: starting with $v^{(0)} \in {\cal C}$, inductively define
\[
v^{(\ell+1)} = \arg \min ~ g(v^{(\ell)}) + \nabla g(v^{(\ell)}) \cdot (v -
v^{(\ell)}) \quad \textrm{subject to} \quad v \in {\cal C}.
\]
Each iterate is now the solution to a convex optimization problem. In
the case \eqref{eq:logsum2} of interest, this gives
\[
(x^{(\ell+1)}, u^{(\ell+1)}) = \arg \min ~ \sum_{i = 1}^\ncol
\frac{u_i}{u^{(\ell)}_i + \epsilon}
\quad \textrm{subject to} \quad \begin{array}{l} y  = \Phi \xhat, \\
  |\xhat_i| \le u_i, \,\,\, i = 1, \ldots, \ncol,
\end{array}
\]
which is of course equivalent to
\[
x^{(\ell+1)} = \arg \min ~ \sum_{i = 1}^\ncol
\frac{|x_i|}{|x^{(\ell)}_i|+ \epsilon} \quad \textrm{subject to}
\quad y = \Phi \xhat.
\]
One now recognizes our iterative algorithm.

In two papers~\cite{loboPortfolio,fazelRank}, Fazel et al.\ have
considered the same reweighted $\ell_1$ minimization algorithm as in
Section~\ref{sec:alg}, first as a heuristic algorithm for
applications in portfolio optimization~\cite{loboPortfolio}, and
second as a special case of an iterative algorithm for minimizing
the rank of a matrix subject to convex constraints~\cite{fazelRank}.
Using general theory, they argue that $\sum_{i = 1}^\ncol
\log(|x_i^{(\ell)}| + \epsilon)$ converges to a local minimum of
$g(x) = \sum_{i = 1}^\ncol \log(|x_i| + \epsilon)$ (note that this
not saying that the sequence $(x^{(\ell)})$ converges). Because the
log-sum penalty function is concave, one cannot expect this
algorithm to always find a global minimum. As a result, it is
important to choose a suitable starting point for the algorithm.
Like~\cite{fazelRank}, we have suggested initializing with the
solution to ($\Pone$), the unweighted $\ell_1$ minimization. In
practice we have found this to be an effective strategy. Further
connections between our work and FOCUSS strategies are discussed at
the end of the paper.

\begin{figure}
\begin{center}
\includegraphics[scale=0.5333]{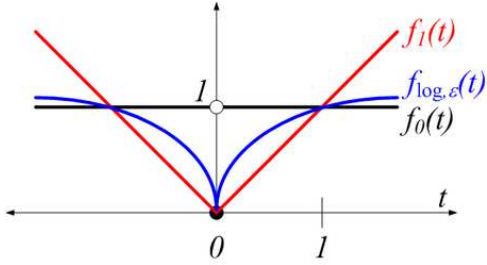}
\end{center}
\caption{\small\sl At the origin, the canonical
$\ell_0$ sparsity count $f_0(t)$ is better approximated by the
log-sum penalty function $f_{\log,\epsilon}(t)$ than by the traditional convex
$\ell_1$ relaxation $f_1(t)$.}\label{fig:ell0ell1log}
\end{figure}

The connection with the log-sum penalty function provides a basis
for understanding why reweighted $\ell_1$ minimization can improve
the recovery of sparse signals. In particular, the log-sum penalty
function has the potential to be much more sparsity-encouraging than
the $\ell_1$ norm. Consider, for example, three potential penalty
functions for scalar magnitudes $t$:
\[
f_0(t) = 1_{\{t \neq 0\}}, \quad f_1(t) = |t|, \quad \textrm{and} \quad
f_{\textrm{log},\epsilon}(t) \propto \log(1 + |t|/\epsilon),
\]
where the constant of proportionality is set such that
$f_{\textrm{log},\epsilon}(1) = 1 = f_0(1) = f_1(1)$, see
Figure~\ref{fig:ell0ell1log}. The first ($\ell_0$-like) penalty
function $f_0$ has infinite slope at $t = 0$, while its convex
($\ell_1$-like) relaxation $f_1$ has unit slope at the origin. The
concave penalty function $f_{\textrm{log},\epsilon}$, however, has
slope at the origin that grows roughly as $1/\epsilon$ when $\epsilon
\rightarrow 0$. Like the $\ell_0$ norm, this allows a relatively large
penalty to be placed on small nonzero coefficients and more strongly
encourages them to be set to zero. In fact,
$f_{\textrm{log},\epsilon}(t)$ tends to $f_0(t)$ as $\epsilon
\rightarrow 0$. Following this argument, it would appear that
$\epsilon$ should be set arbitrarily small, to most closely make the
log-sum penalty resemble the $\ell_0$ norm.  Unfortunately, as
$\epsilon \rightarrow 0$, it becomes more likely that the iterative
reweighted $\ell_1$ algorithm will get stuck in an undesirable local
minimum. As shown in Section~\ref{sec:exp}, a cautious choice of
$\epsilon$ (slightly smaller than the expected nonzero magnitudes of
$x$) provides the stability necessary to correct for inaccurate
coefficient estimates while still improving upon the unweighted
$\ell_1$ algorithm for sparse recovery.

\subsection{Variations}
\label{sec:variations}

One could imagine a variety of possible reweighting functions in place
of \eqref{eq:rwrule}. We have experimented with alternatives,
including a binary (large/small) setting of $w_i$ depending on the
current guess. Though such alternatives occasionally provide superior
reconstruction of sparse signals, we have found the rule
\eqref{eq:rwrule} to perform well in a variety of experiments and
applications.

Alternatively, one can attempt to minimize a concave
function other than the log-sum penalty. For instance, we may consider
\[
g(x) = \sum_{i = 1}^n \textrm{atan}(|x_i|/\epsilon)
\]
in lieu of $\sum_{i = 1}^n \log(1 + |x_i|/\epsilon)$.  The function
$\textrm{atan}$ is bounded above and $\ell_0$-like. If $\xhat$ is the
current guess, this proposal updates the sequence of weights as $w_i =
1/(x_i^2 + \epsilon^2)$. There are of course many possibilities of
this nature and they tend to work well (sometimes better than the
log-sum penalty). Because of space limitations, however, we will limit
ourselves to empirical studies of the performance of the log-sum
penalty, and leave the choice of other penalties for further research.

\subsection{Historical progression}
\label{sec:irls}

The development of the reweighted $\ell_1$ algorithm has an
interesting historical parallel with the use of Iteratively Reweighted
Least Squares (IRLS) for robust statistical
estimation~\cite{schlossmacher73it,holland77ro,huber81ro}. Consider a
regression problem $Ax = b$ where the observation matrix $A$ is
overdetermined. It was noticed that standard least squares regression,
in which one minimizes $\norm{r}_2$ where $r = Ax - b$ is the residual
vector, lacked robustness vis a vis outliers. To defend against this,
IRLS was proposed as an iterative method to minimize instead the
objective
$$
\min_x \, \sum_i \rho (r_i(x)),
$$
where $\rho(\cdot)$ is a penalty function such as the $\ell_1$
norm~\cite{schlossmacher73it,yarlagadda85fa}. This minimization can be
accomplished by solving a sequence of weighted least-squares problems
where the weights $\{w_i\}$ depend on the previous residual $w_i =
\rho'(r_i)/r_i$.  For typical choices of $\rho$ this dependence is in
fact inversely proportional---large residuals will be penalized less
in the subsequent iteration and vice versa---as is the case with our
reweighted $\ell_1$ algorithm.  Interestingly, just as IRLS involved
iteratively reweighting the $\ell_2$-norm in order to better
approximate an $\ell_1$-like criterion, our algorithm involves
iteratively reweighting the $\ell_1$-norm in order to better
approximate an $\ell_0$-like criterion.

\section{Numerical experiments}
\label{sec:exp}

We present a series of experiments demonstrating the benefits of
reweighting the $\ell_1$ penalty. We will see that the requisite
number of measurements to recover or approximate a signal is
typically reduced, in some cases by a substantial amount. We also
demonstrate that the reweighting approach is robust and broadly
applicable, providing examples of sparse and compressible signal
recovery, noise-aware recovery, model selection, error correction,
and 2-dimensional total-variation minimization. Meanwhile, we
address important issues such as how one can choose $\epsilon$
wisely and how robust is the algorithm to this choice, and how many
reweighting iterations are needed for convergence.

\subsection{Sparse signal recovery}
\label{sec:sparse1}

The purpose of this first experiment is to demonstrate (1) that
reweighting reduces the necessary sampling rate for sparse signals
(2) that this recovery is robust with respect to the choice of
$\epsilon$ and (3) that few reweighting iterations are typically
needed in practice. The setup for each trial is as follows.  We
select a sparse signal $\true$ of length $\ncol = 256$ with
$\|\true\|_{\ell_0} = \nsparse$. The $\nsparse$ nonzero spike
positions are chosen randomly, and the nonzero values are chosen
randomly from a zero-mean unit-variance Gaussian distribution.  We
set $\nrow = 100$ and sample a random $\nrow \times \ncol$ matrix
$\Phi$ with i.i.d.~Gaussian entries, giving the data $y = \Phi
\true$.  To recover the signal, we run several reweighting
iterations with equality constraints (see Section~\ref{sec:alg}).
The parameter $\epsilon$ remains fixed during these iterations.
Finally, we run 500 trials for various fixed combinations of
$\nsparse$ and $\epsilon$.

\begin{figure}
\begin{center}
\begin{tabular}{cc}
~~ \includegraphics[scale=0.38]{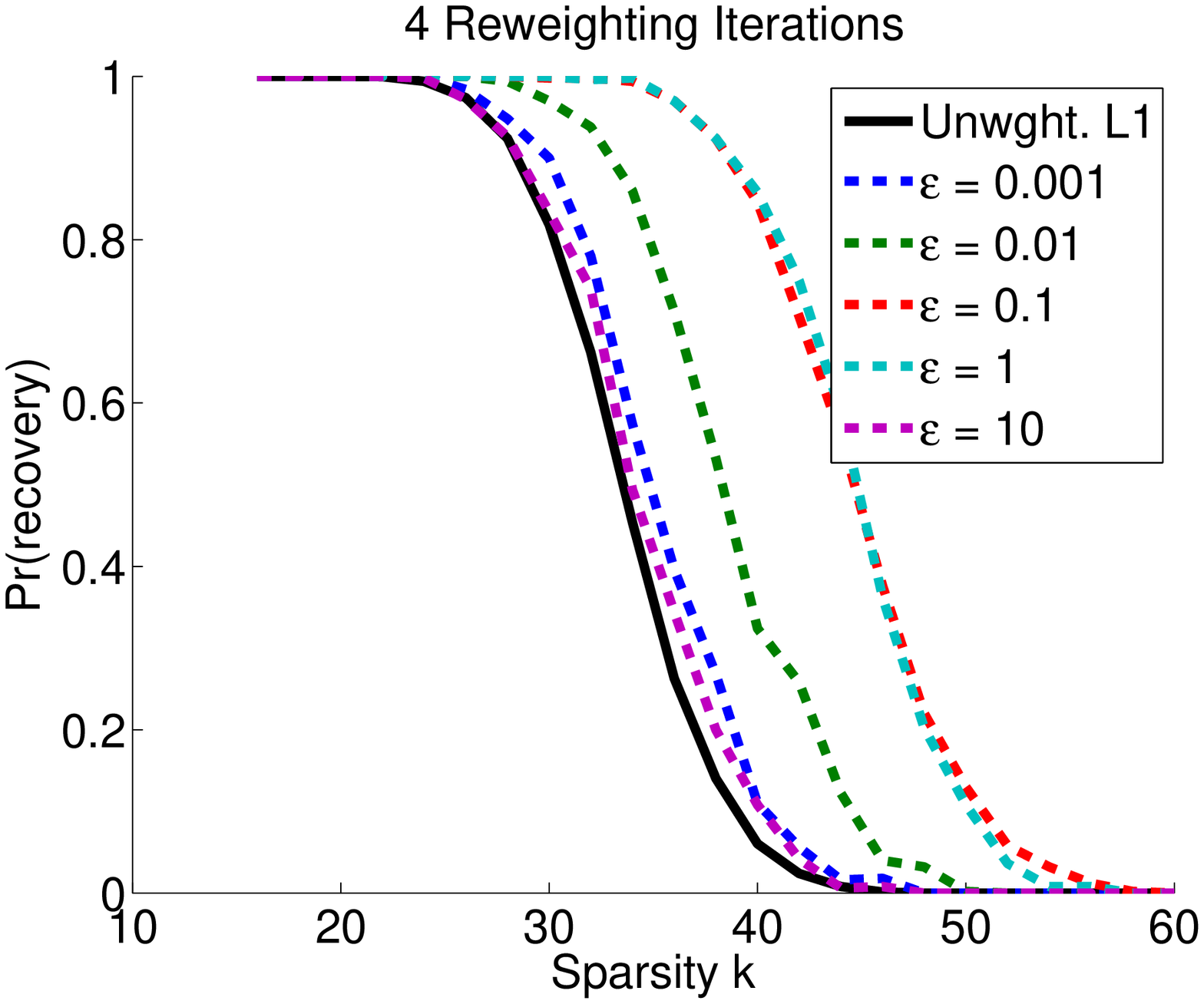} ~~ &
~~ \includegraphics[scale=0.38]{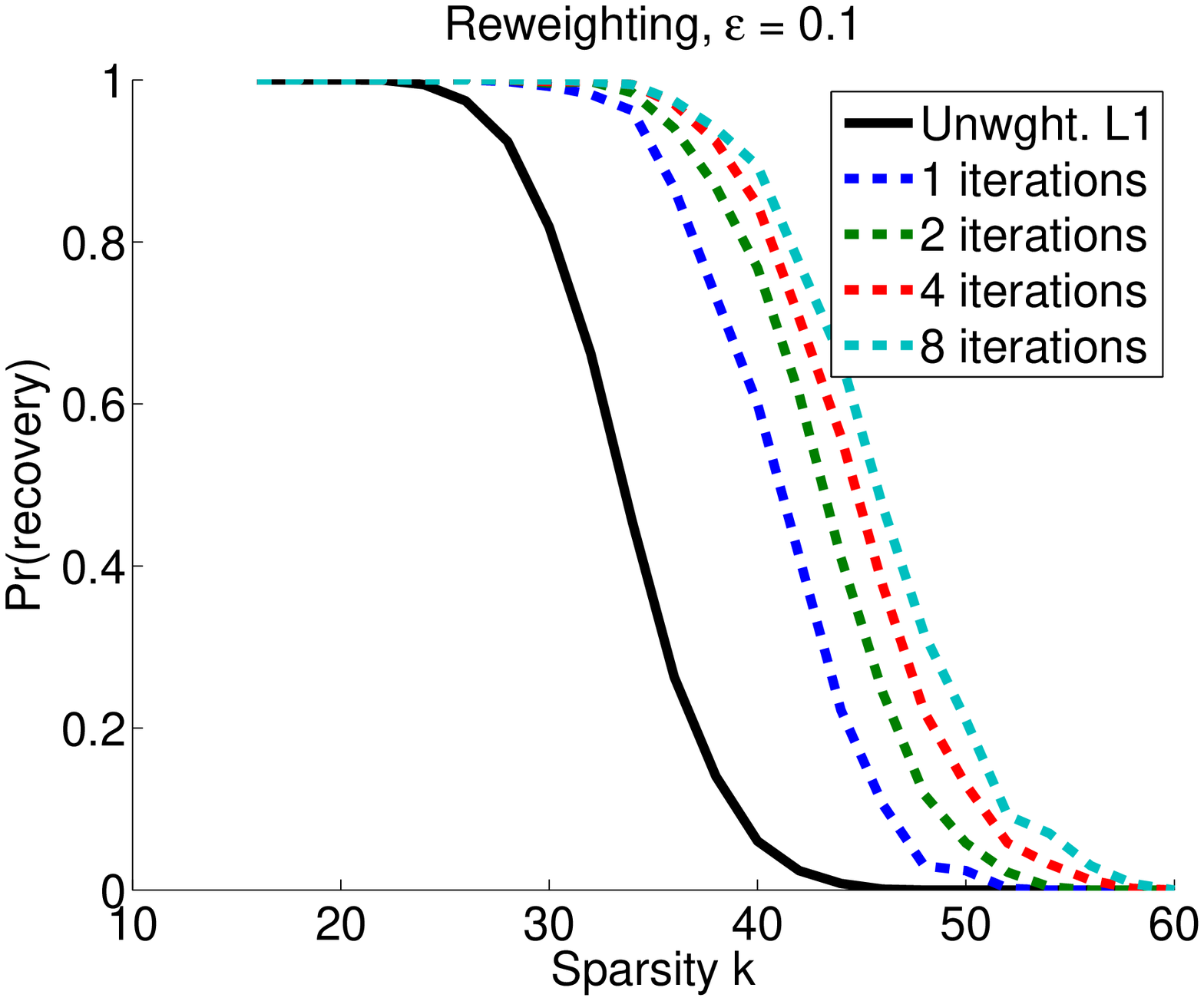} ~~\\
(a) & (b)
\end{tabular}
\end{center}
\caption{\small\sl Sparse signal recovery from $\nrow=100$ random
  measurements of a length $\ncol = 256$ signal. The probability of
  successful recovery depends on the sparsity level $\nsparse$. The
  dashed curves represent a reweighted $\ell_1$ algorithm that
  outperforms the traditional unweighted $\ell_1$ approach (solid
  curve). (a) Performance after 4 reweighting iterations as a function
  of $\epsilon$. (b) Performance with fixed $\epsilon = 0.1$ as a
  function of the number of reweighting iterations.
}\label{fig:sparse1}
\end{figure}

Figure~\ref{fig:sparse1}(a) compares the performance of unweighted
$\ell_1$ to reweighted $\ell_1$ for various values of the parameter
$\epsilon$. The solid line plots the probability of perfect signal
recovery (declared when $\|\true-x\|_{\ell_\infty} \le 10^{-3}$) for
the unweighted $\ell_1$ algorithm as a function of the sparsity
level $\nsparse$. The dashed curves represent the performance after
4 reweighted iterations for several different values of the
parameter $\epsilon$. We see a marked improvement over the
unweighted $\ell_1$ algorithm; with the proper choice of $\epsilon$,
the requisite oversampling factor $\nrow/\nsparse$ for perfect
signal recovery has dropped from approximately $100/25 = 4$ to
approximately $100/33 \approx 3$. This improvement is also fairly
robust with respect to the choice of $\epsilon$, with a suitable
rule being about $10\%$ of the standard deviation of the nonzero
signal coefficients.

Figure~\ref{fig:sparse1}(b) shows the performance, with a fixed
value of $\epsilon = 0.1$, of the reweighting algorithm for various
numbers of reweighted iterations. We see that much of the benefit
comes from the first few reweighting iterations, and so the added
computational cost for improved signal recovery is quite moderate.

\subsection{Sparse and compressible signal recovery with adaptive
  choice of $\epsilon$}

We would like to confirm the benefits of reweighted $\ell_1$
minimization for compressible signal recovery and consider the
situation when the parameter $\epsilon$ is not provided in advance
and must be estimated during reconstruction.  We propose an
experiment in which each trial is designed as follows.  We sample a
signal of length $\ncol = 256$ from one of three types of
distribution: (1)~$\nsparse$-sparse with i.i.d.~Gaussian entries,
(2)~$\nsparse$-sparse with i.i.d.~symmetric Bernoulli $\pm 1$
entries, or (3)~compressible, constructed by randomly permuting the
sequence $\{i^{-1/p}\}_{i = 1}^\ncol$ for a fixed $p$, applying
random sign flips, and normalizing so that $\|\true\|_{\ell_\infty}
= 1$.  We set $\nrow = 128$ and sample a random $\nrow \times \ncol$
matrix $\Phi$ with i.i.d.~Gaussian entries. To recover the signal,
we again solve a reweighted $\ell_1$ minimization with equality
constraints $y = \Phi \true = \Phi \xhat$.  In this case, however,
we adapt $\epsilon$ at each iteration as a function of the current
guess $x^{(\ell)}$; step 3 of the algorithm is modified as follows:
\begin{enumerate}
\setcounter{enumi}{2}
\item Let $(|x|_{(i)})$ denote a reordering of $(|\xhat_i|)$ in
  decreasing order of magnitude. Set
\[
\epsilon =
\max\left\{|x^{(\ell)}|_{(i_0)}, 10^{-3}\right\},
\]
where $i_0 = {\nrow}/[4\log(\ncol/\nrow)]$.  Define $w^{(\ell+1)}$
as in \eqref{eq:rwrule}.
\end{enumerate}
Our motivation for choosing this value for $\epsilon$ is based on
the anticipated accuracy of $\ell_1$ minimization for arbitrary
signal recovery. In general, the reconstruction quality afforded by
$\ell_1$ minimization is comparable (approximately) to the best
$i_0$-term approximation to $\true$, and so we expect
approximately this many signal components to be approximately
correct. Choosing the smallest of these gives us a rule of thumb for
choosing $\epsilon$.

\begin{figure}
\begin{center}
\begin{tabular}{ccc}
\includegraphics[scale=0.26]{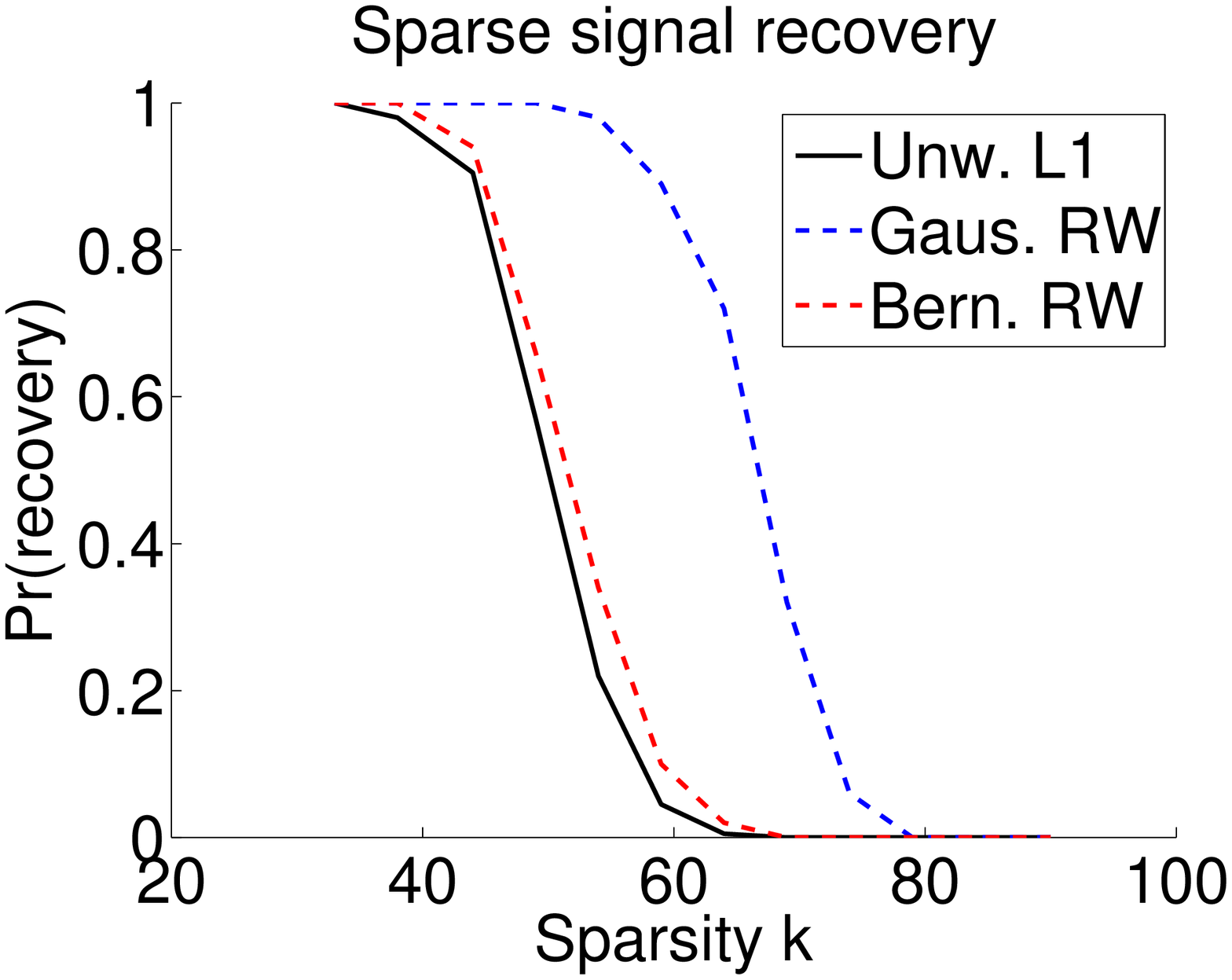} &
\includegraphics[scale=0.26]{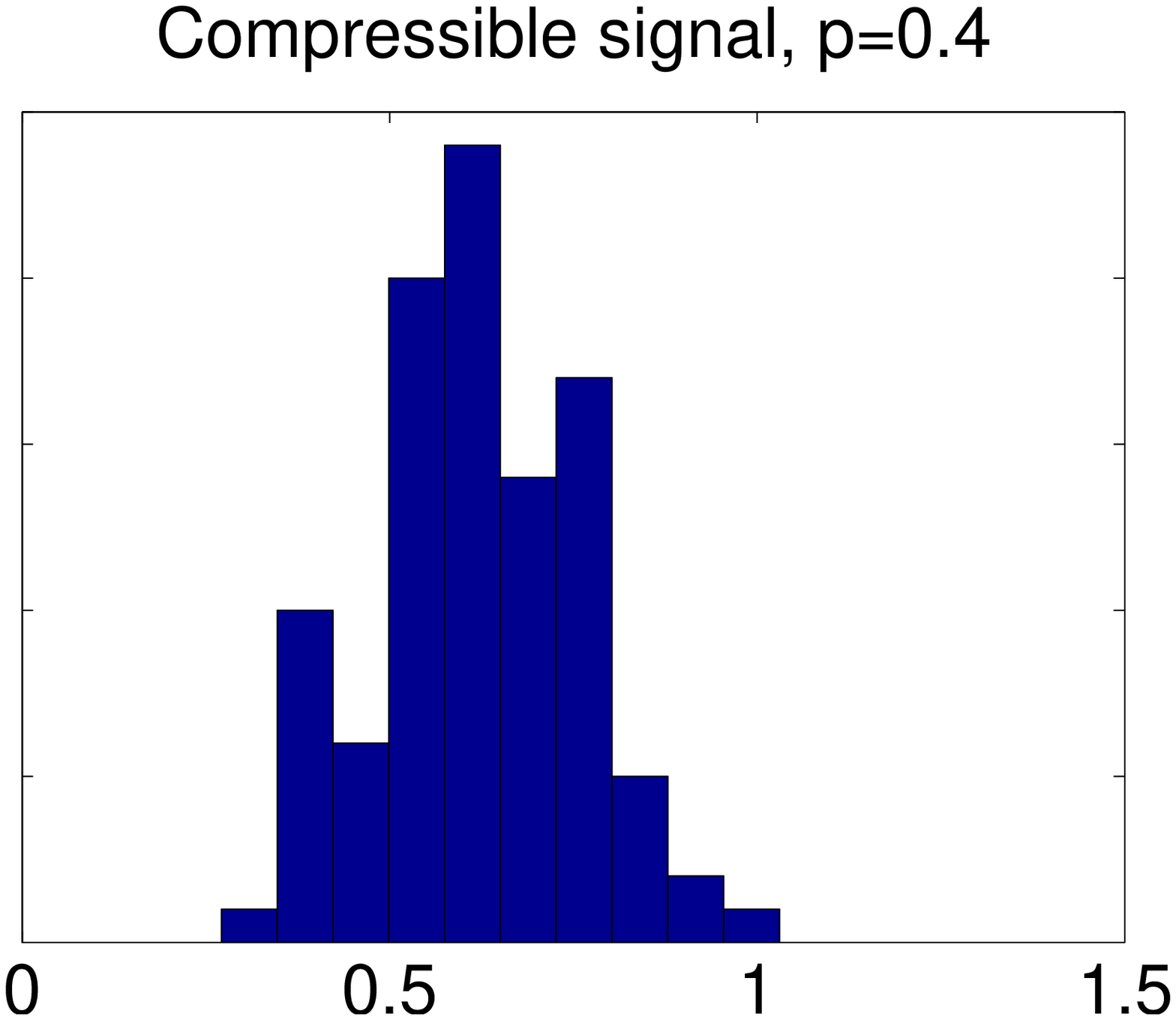} &
\includegraphics[scale=0.26]{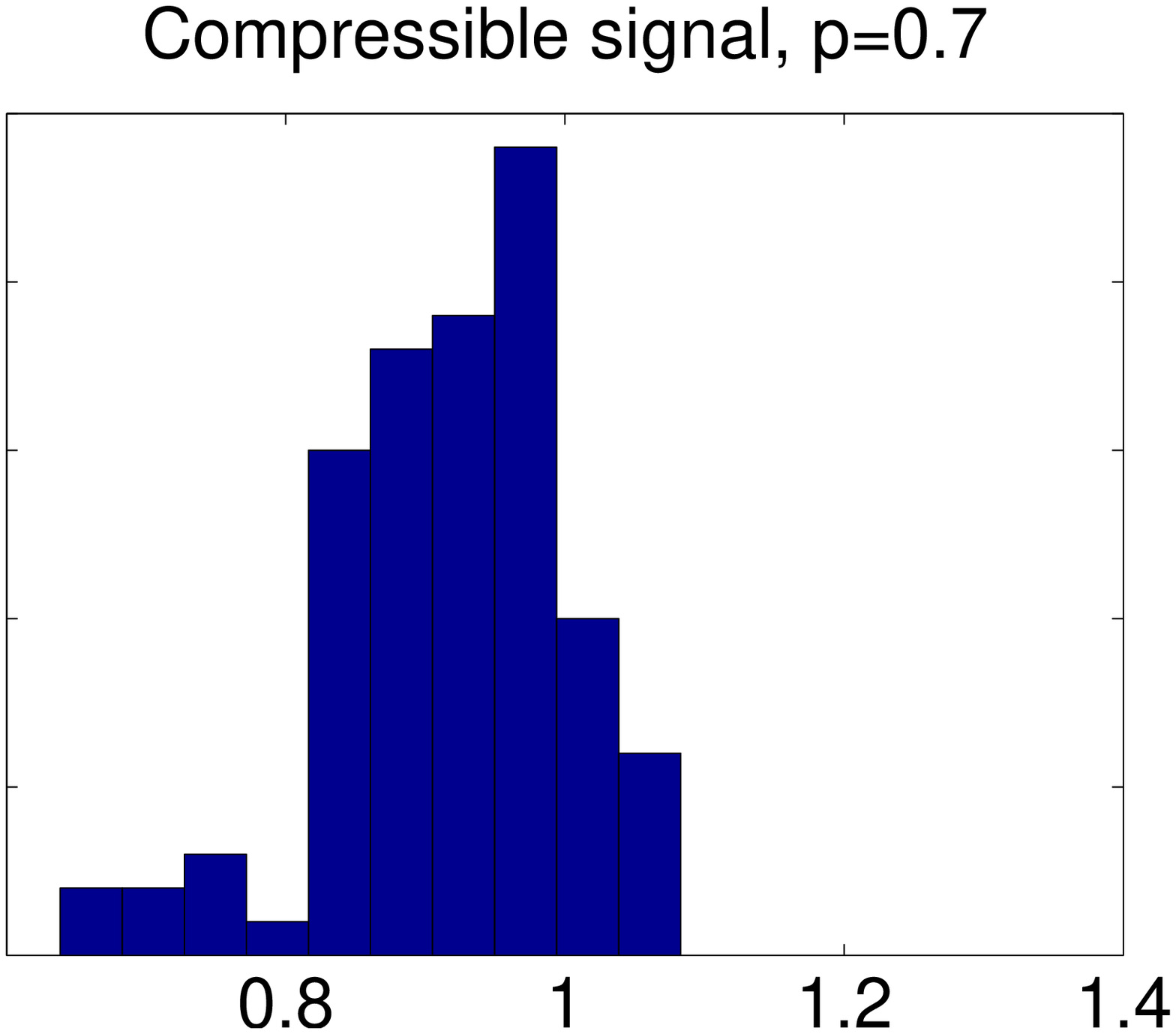}\\
(a) & (b) & (c)
\end{tabular}
\end{center}
\caption{\small\sl (a) Improvements in sparse signal
recovery from reweighted $\ell_1$ minimization when compared to
unweighted $\ell_1$ minimization (solid black curve). The dashed
blue curve corresponds to sparse signals with Gaussian coefficients;
the dashed red curve corresponds to sparse signals with Bernoulli
coefficients. (b),(c) Improvements in compressible signal recovery
from reweighted $\ell_1$ minimization when compared to unweighted
$\ell_1$ minimization; signal coefficients decay as $n^{-1/p}$ with
(b) $p=0.4$ and (c) $p=0.7$. Histograms indicate the $\ell_2$
reconstruction error improvements afforded by the reweighted
algorithm. }\label{fig:adaptive}
\end{figure}

We run 100 trials of the above experiment for each signal type. The
results for the $\nsparse$-sparse experiments are shown in
Figure~\ref{fig:adaptive}(a). The solid black line indicates the
performance of unweighted $\ell_1$ recovery (success is declared
when $\|\true-x\|_{\ell_\infty} \le 10^{-3}$). This curve is the
same for both the Gaussian and Bernoulli coefficients, as the
success or failure of unweighted $\ell_1$ minimization depends only
on the support and sign pattern of the original sparse signal. The
dashed curves indicate the performance of reweighted $\ell_1$
minimization for Gaussian coefficients (blue curve) and Bernoulli
coefficients (red curve) with $\ell_{\mathrm{max}} = 4$. We see a
substantial improvement for recovering sparse signals with Gaussian
coefficients, yet we see only very slight improvement for recovering
sparse signals with Bernoulli coefficients. This discrepancy likely
occurs because the decay in the sparse Gaussian coefficients allows
large coefficients to be easily identified and significantly
downweighted early in the reweighting algorithm. With Bernoulli
coefficients there is no such ``low-hanging fruit''.

The results for compressible signals are shown in
Figure~\ref{fig:adaptive}(b),(c). Each plot represents a histogram,
over 100 trials, of the $\ell_2$ reconstruction error improvement
afforded by reweighting, namely, $\|\true -
x^{(4)}\|_{\ell_2}/\|\true-\xhat^{(0)}\|_{\ell_2}$. We see the greatest
improvements for smaller $p$ corresponding to sparser signals, with
reductions in $\ell_2$ reconstruction error up to $50\%$ or more. As
$p \rightarrow 1$, the improvements diminish.

\subsection{Recovery from noisy measurements}
\label{sec:qcl1}

Reweighting can be applied to a noise-aware version of $\ell_1$
minimization, further improving the recovery of signals from noisy
data. We observe $y = \Phi \true + z$, where $z$ is a noise term which
is either stochastic or deterministic. To recover $\true$, we adapt
quadratically-constrained $\ell_1$
minimization~\cite{tibshirani96re,CandesSSR}, and modify step 2 of the
reweighted $\ell_1$ algorithm with equality constraints (see
Section~\ref{sec:alg}) as
\begin{equation}
  \xhat^{(\ell)} = \arg\min \|W^{(\ell)} \xhat\|_{\ell_1} \quad \textrm{subject to} \quad \|y - \Phi
  \xhat\|_{\ell_2} \le \delta. \label{eq:rwqcl1}
\end{equation}
The parameter $\delta$ is adjusted so that the true vector $\true$
be feasible (resp.~feasible with high probability) for
\eqref{eq:rwqcl1} in the case where $z$ is deterministic
(resp.~stochastic).

To demonstrate how this proposal improves on plain $\ell_1$
minimization, we sample a vector of length $\ncol = 256$ from one of
three types of distribution: (1)~$\nsparse$-sparse with $\nsparse=38$
and i.i.d.~Gaussian entries, (2)~$\nsparse$-sparse with $\nsparse=38$
and i.i.d.~symmetric Bernoulli $\pm 1$ entries, or (3)~compressible,
constructed by randomly permuting the sequence $\{i^{-1/p}\}_{i =
  1}^\ncol$ for a fixed $p$, applying random sign flips, and
normalizing so that $\|\true\|_{\ell_\infty} = 1$. The matrix $\Phi$
is $128 \times 256$ with i.i.d.~Gaussian entries whose columns are
subsequently normalized, and the noise vector $z$ is drawn from an
i.i.d.~Gaussian zero-mean distribution properly rescaled so that
$\|z\|_{\ell_2} = \beta \|\Phi x\|_{\ell_2}$ with $\beta = 0.2$; i.e.,
$z = \sigma z_0$ where $z_0$ is standard white noise and $\sigma =
\beta \|\Phi x\|_{\ell_2}/\|z_0\|_{\ell_2}$.  The parameter $\delta$
is set to $\delta^2 = \sigma^2 (\nrow + 2\sqrt{2\nrow})$ as this
provides a likely upper bound on $\|z\|_{\ell_2}$.  We set $\epsilon$
to be the empirical maximum value of $\|\Phi^\ast \xi\|_{\ell_\infty}$
over several realizations of a random vector $\xi \sim
\gaussian(0,\sigma^2 I_\nrow)$. (This gives a rough estimate for the
noise amplitude in the signal domain, and hence, a baseline above
which significant signal components could be identified.)

\begin{figure}
\begin{center}
\begin{tabular}{ccc}
\includegraphics[scale=0.26]{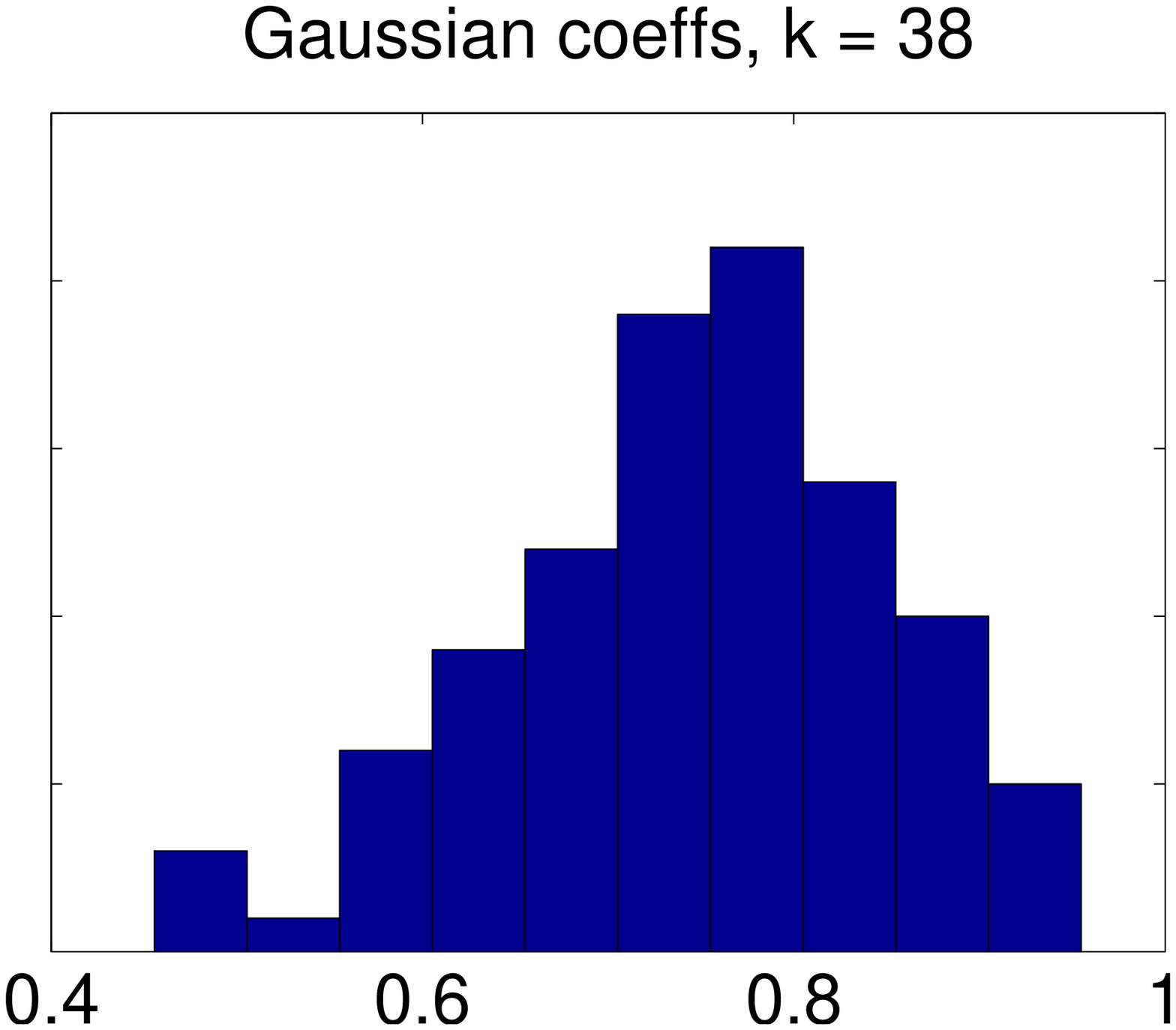} &
\includegraphics[scale=0.26]{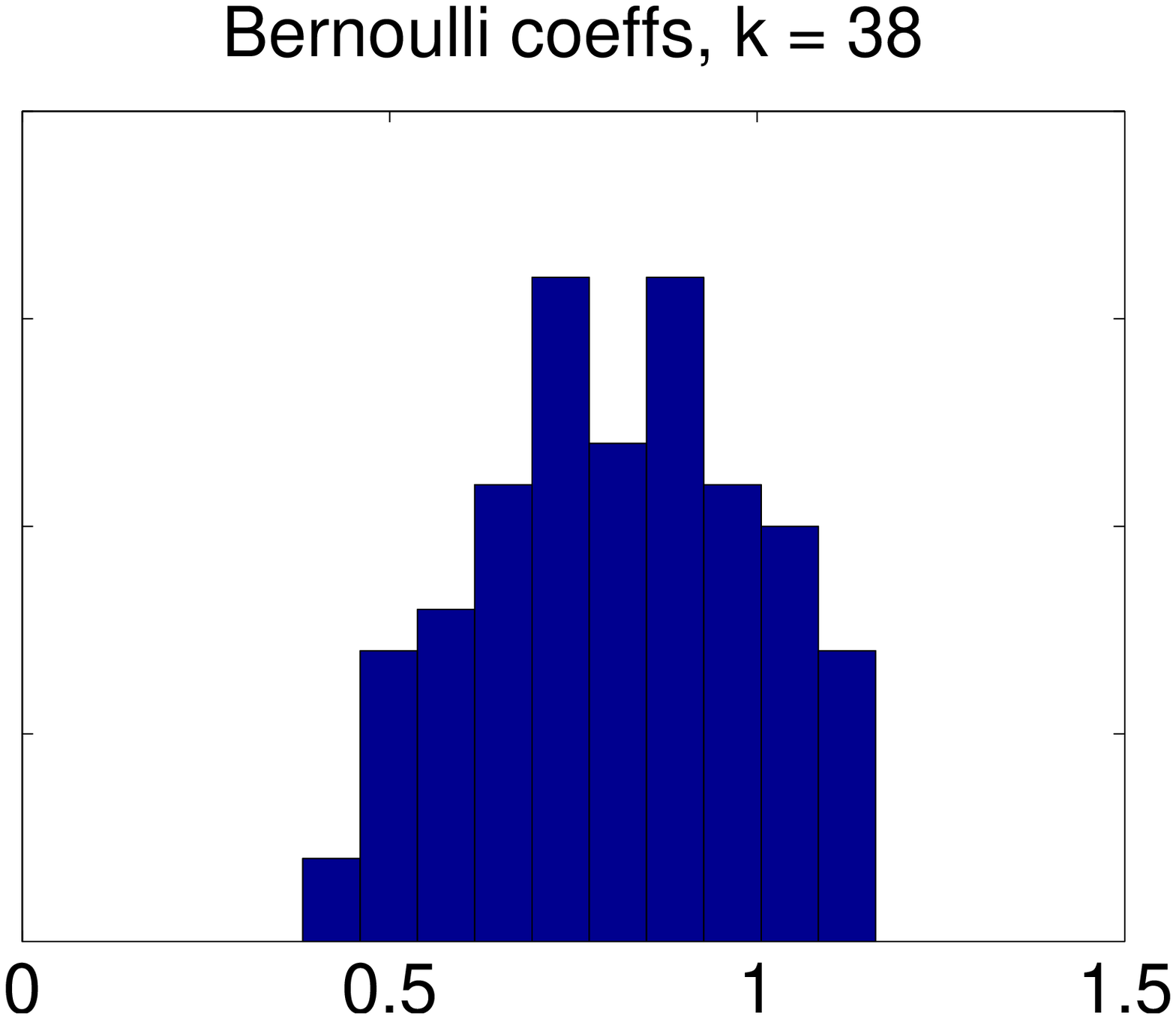} &
\includegraphics[scale=0.26]{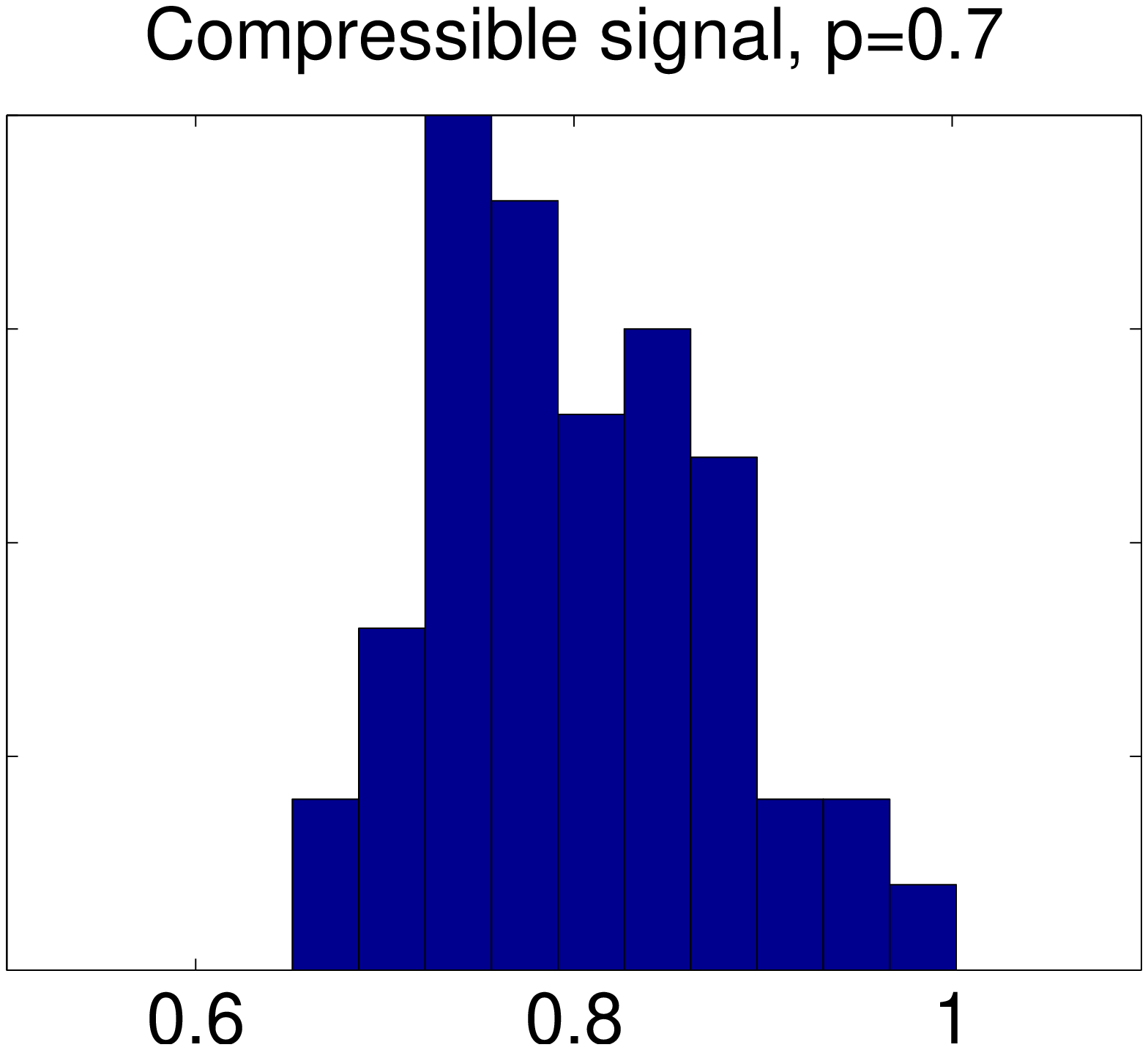}\\
(a) & (b) & (c)
\end{tabular}
\end{center}
\caption{\small\sl Sparse and compressible signal
reconstruction from noisy measurements. Histograms indicate the
$\ell_2$ reconstruction error improvements afforded by the
reweighted quadratically-constrained $\ell_1$ minimization for
various signal types.}\label{fig:noise1}
\end{figure}

We run 100 trials for each signal type. Figure~\ref{fig:noise1} shows
histograms of the $\ell_2$ reconstruction error improvement afforded
by 9 iterations, i.e., each histogram documents $\|\true
-\xhat^{(9)}\|_{\ell_2}/\|\true - \xhat^{(0)}\|_{\ell_2}$ over 100
trials. We see in these experiments that the reweighted
quadratically-constrained $\ell_1$ minimization typically offers
improvements $\|\true -\xhat^{(9)}\|_{\ell_2}/\|\true -
\xhat^{(0)}\|_{\ell_2}$ in the range $0.5-1$ in many examples. The
results for sparse Gaussian spikes are slightly better than for sparse
Bernoulli spikes, though both are generally favorable. Similar
behavior holds for compressible signals, and we have observed that
smaller values of $p$ (sparser signals) allow the most improvement.

\subsection{Statistical estimation}

Reweighting also enhances statistical estimation as well.  Suppose
we observe $y = \Phi \true + z$, where $\Phi$ is $\nrow \times
\ncol$ with $\nrow \le \ncol$, and $z$ is a noise vector $z \sim
\mathcal{N}(0,\sigma^2 I_\nrow)$ drawn from an i.i.d.~Gaussian
zero-mean distribution, say. To estimate $\true$, we adapt the
Dantzig selector~\cite{candesDS} and modify step 2 of the reweighted
$\ell_1$ algorithm as
\begin{equation}
  \xhat^{(\ell)} = \arg\min \|W^{(\ell)} \xhat\|_{\ell_1} \quad \textrm{subject to} \quad \|\Phi^\ast(y
  - \Phi \xhat)\|_{\ell_\infty} \le \delta.
\end{equation}
Again $\delta$ is a parameter making sure that the true unknown vector
is feasible with high probability.

To judge this proposal, we consider a sequence of experiments in which
$\true$ is of length $\ncol = 256$ with $\nsparse = 8$ nonzero entries
in random positions. The nonzero entries of $\true$ have
i.i.d.~entries according to the model $x_i = s_i(1+|a_i|)$ where the
sign $s_i = \pm 1$ with probability $1/2$ and $a_i \sim
\mathcal{N}(0,1)$. The matrix $\Phi$ is $72 \times 256$ with
i.i.d.~Gaussian entries whose columns are subsequently normalized just
as before.  The noise vector $(z_i)$ has
i.i.d.~$\mathcal{N}(0,\sigma^2)$ components with $\sigma =
1/3\sqrt{\nsparse/\nrow} \approx 0.11$. The parameter $\delta$ is set
to be the empirical maximum value of $\|\Phi^\ast z\|_{\ell_\infty}$
over several realizations of a random vector $z \sim
\gaussian(0,\sigma^2 I_\nrow)$. We set $\epsilon = 0.1$.

After each iteration of the reweighted Dantzig selector, we also
refine our estimate $\xhat^{(\ell)}$ using the Gauss-Dantzig technique
to correct for a systematic bias~\cite{candesDS}. Let $I = \{i :
|\xhat^{(\ell)}_i| > \alpha \cdot \sigma\}$ with $\alpha = 1/4$. Then
  one substitutes $\xhat^{(\ell)}$ with the least squares estimate
  which solves
\[
\min_{x \in \real^\ncol} \, \|y - \Phi x\|_{\ell_2} \quad
\textrm{subject to} \quad x_i = 0, \,\, i \notin I;
\]
that is, by regressing $y$ onto the subset of columns indexed by $I$.

\begin{figure}
\begin{center}
\begin{tabular}{cc}
\includegraphics[scale=0.4]{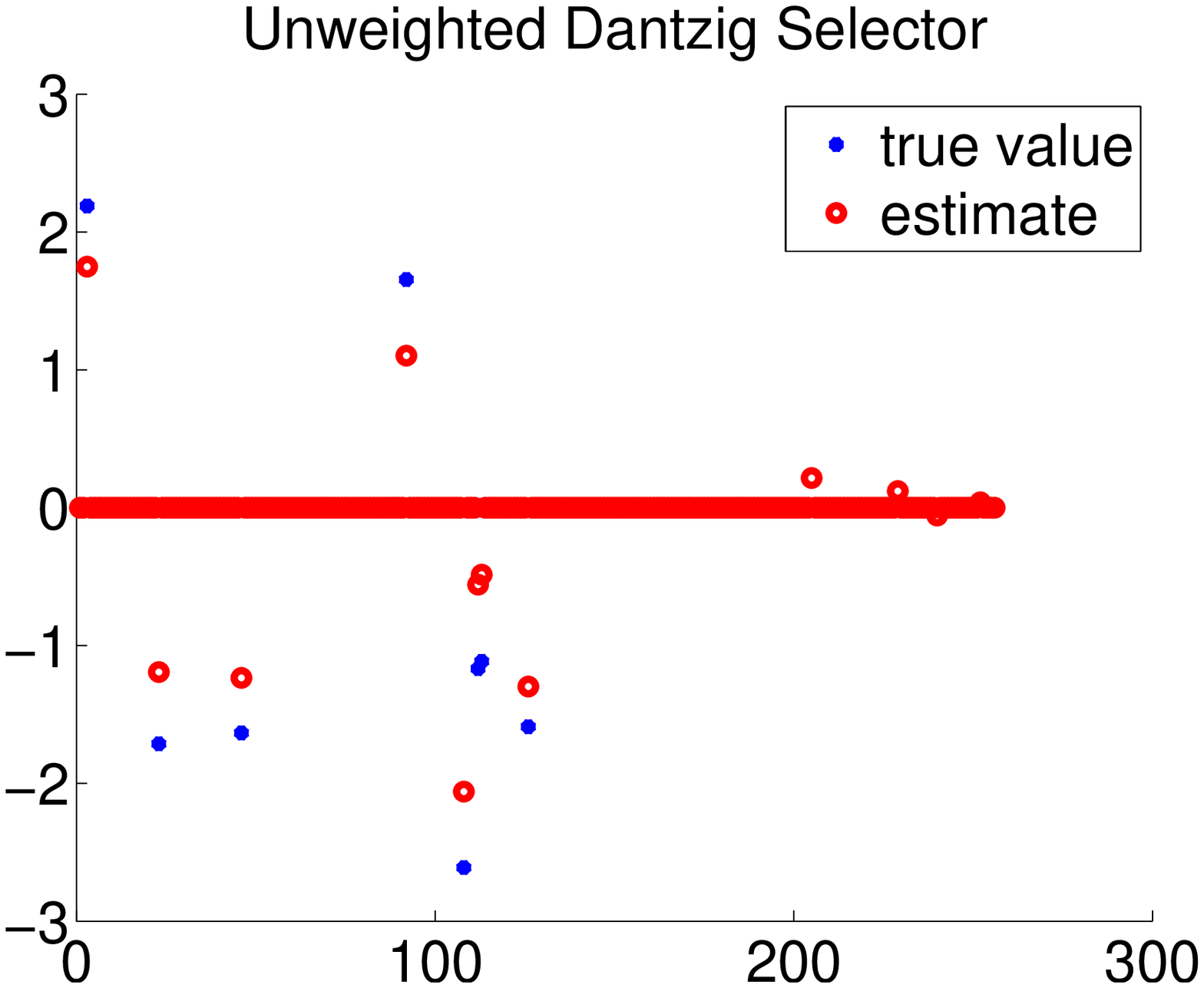} &
\includegraphics[scale=0.4]{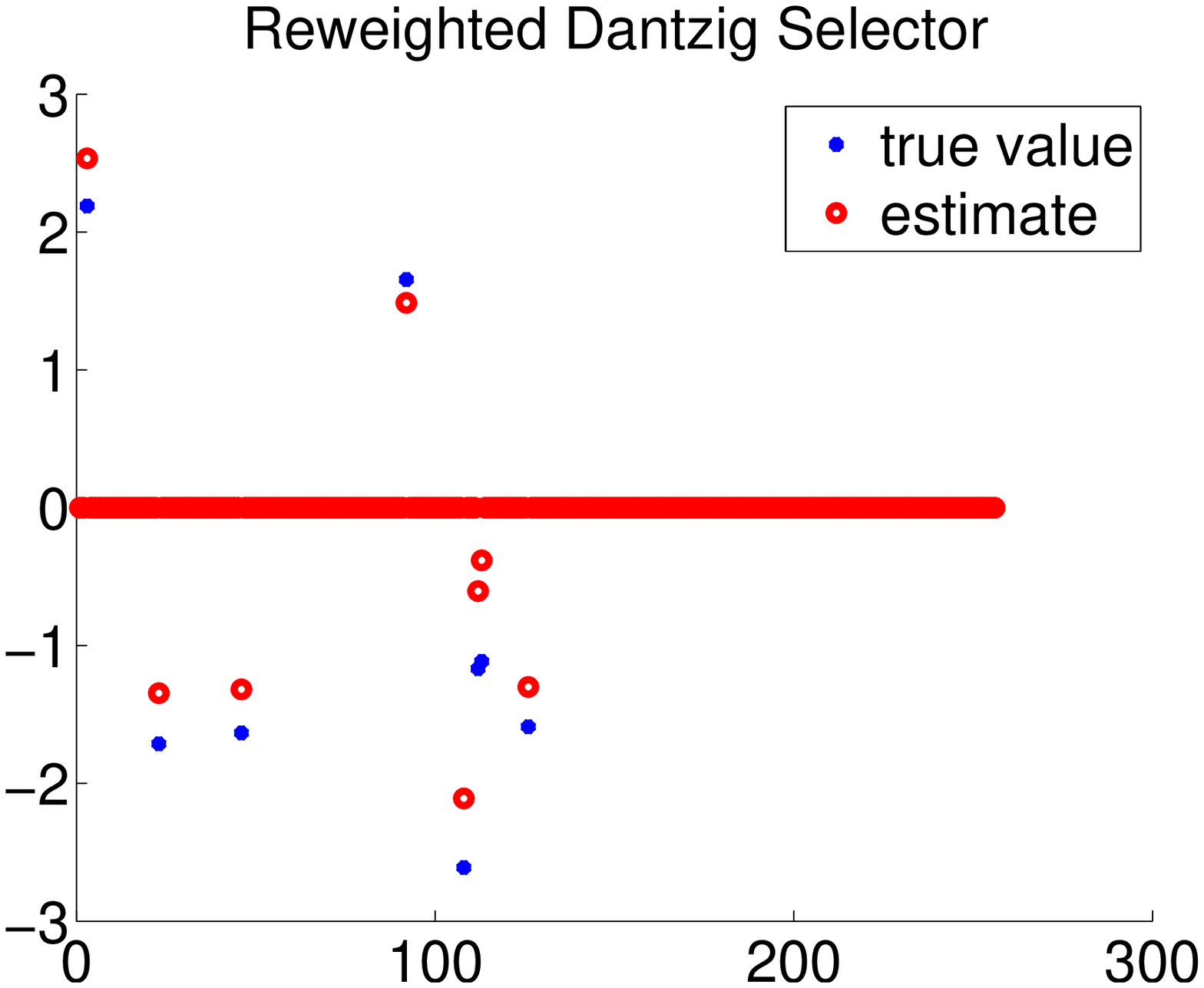}\\
(a) & (b) \\
\includegraphics[scale=0.4]{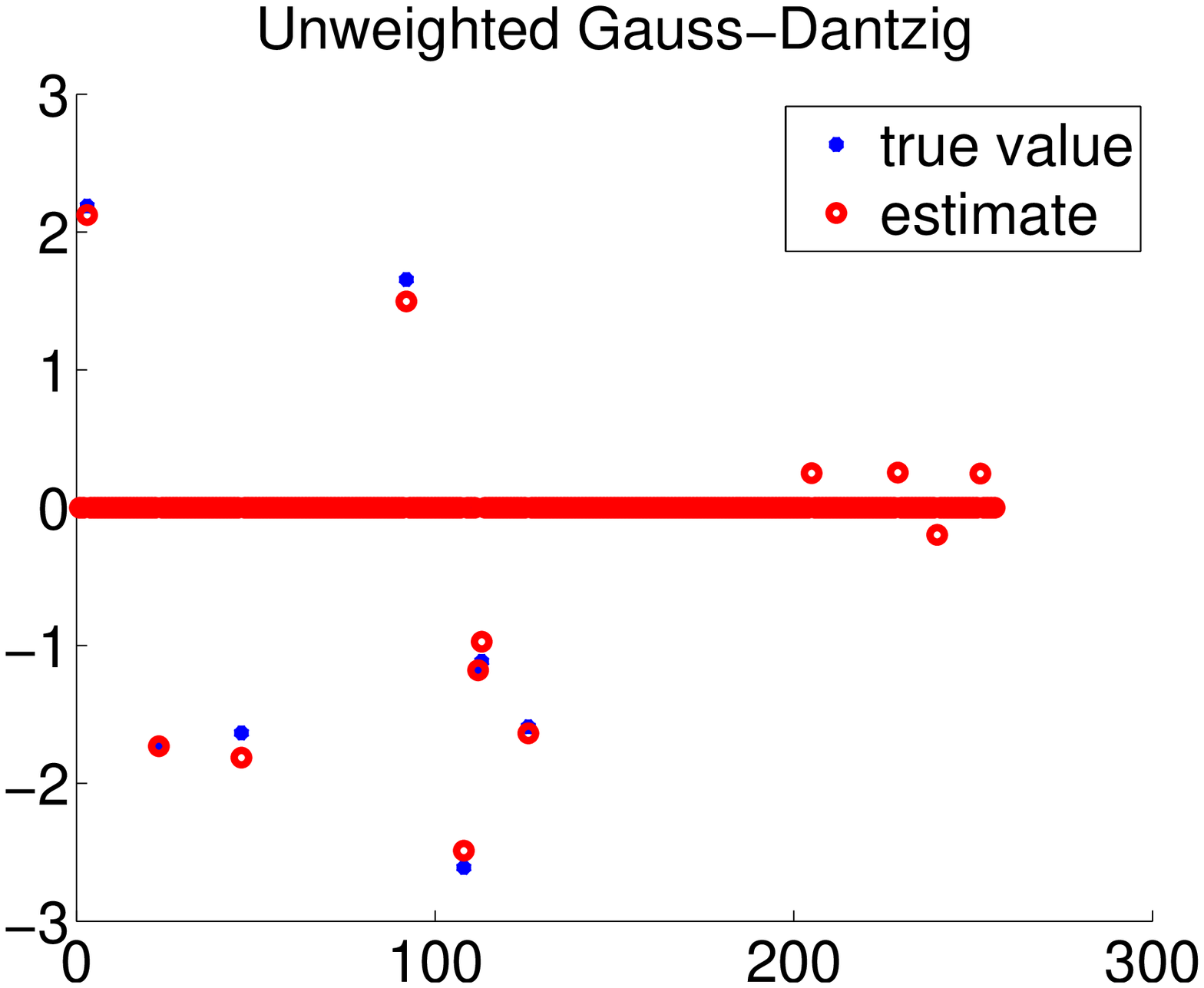} &
\includegraphics[scale=0.4]{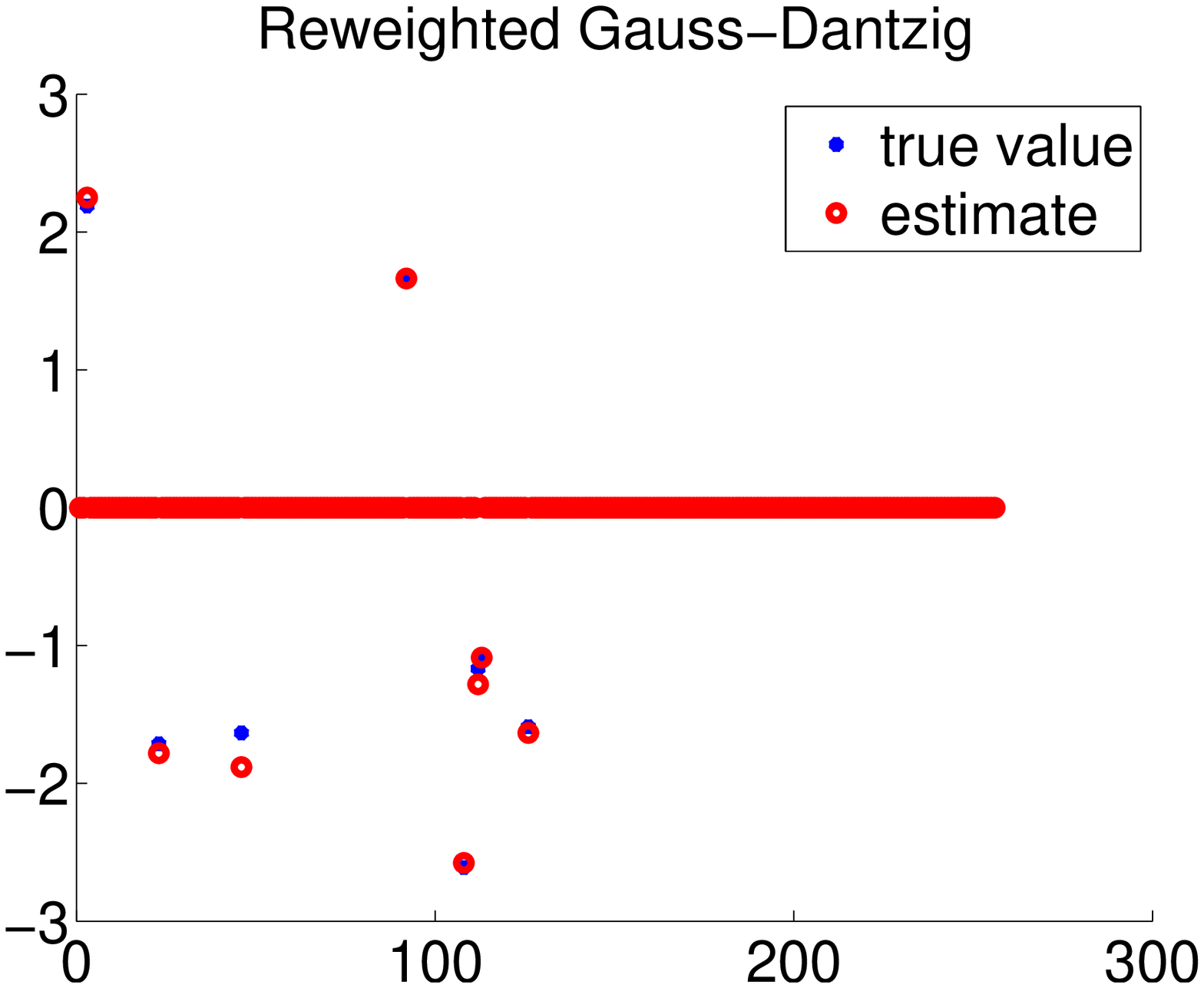} \\
(c) & (d)
\end{tabular}
\end{center}
\caption{\small\sl Reweighting the Dantzig selector.
Blue asterisks indicate the original signal $\true$; red circles
indicate the recovered estimate. (a) Unweighted Dantzig selector.
(b) Reweighted Dantzig selector. (c) Unweighted Gauss-Dantzig
estimate. (d) Reweighted Gauss-Dantzig
estimate.}
\label{fig:rwdz}
\end{figure}

We first report on one trial with $\ell_{\mathrm{max}} = 4$.
Figure~\ref{fig:rwdz}(a) shows the original signal $\true$ along with
the recovery $\xhat^{(0)}$ using the first (unweighted) Dantzig
selector iteration; the error is $\|\true-\xhat^{(0)}\|_{\ell_2} =
1.46$. Figure~\ref{fig:rwdz}(b) shows the Dantzig selector recovery
after $4$ iterations; the error has decreased to $\|\true -
\xhat^{(4)}\|_{\ell_2} = 1.25$. Figure~\ref{fig:rwdz}(c) shows the
Gauss-Dantzig estimate $\xhat^{(0)}$ obtained from the first
(unweighted) Dantzig selector iteration; this decreases the error to
$\|\true-\xhat^{(0)}\|_{\ell_2} = 0.57$.  The estimator correctly
includes all $8$ positions at which $\true$ is nonzero, but also
incorrectly includes $4$ positions at which $\true$ should be zero. In
Figure~\ref{fig:rwdz}(d) we see, however, that all of these mistakes
are rectified in the Gauss-Dantzig estimate $\xhat^{(4)}$ obtained
from the reweighted Dantzig selector; the total error also decreases
to $\|\true - \xhat^{(4)}\|_{\ell_2} = 0.29$.

\begin{figure}
\begin{center}
\begin{tabular}{cc}
\includegraphics[scale=0.4]{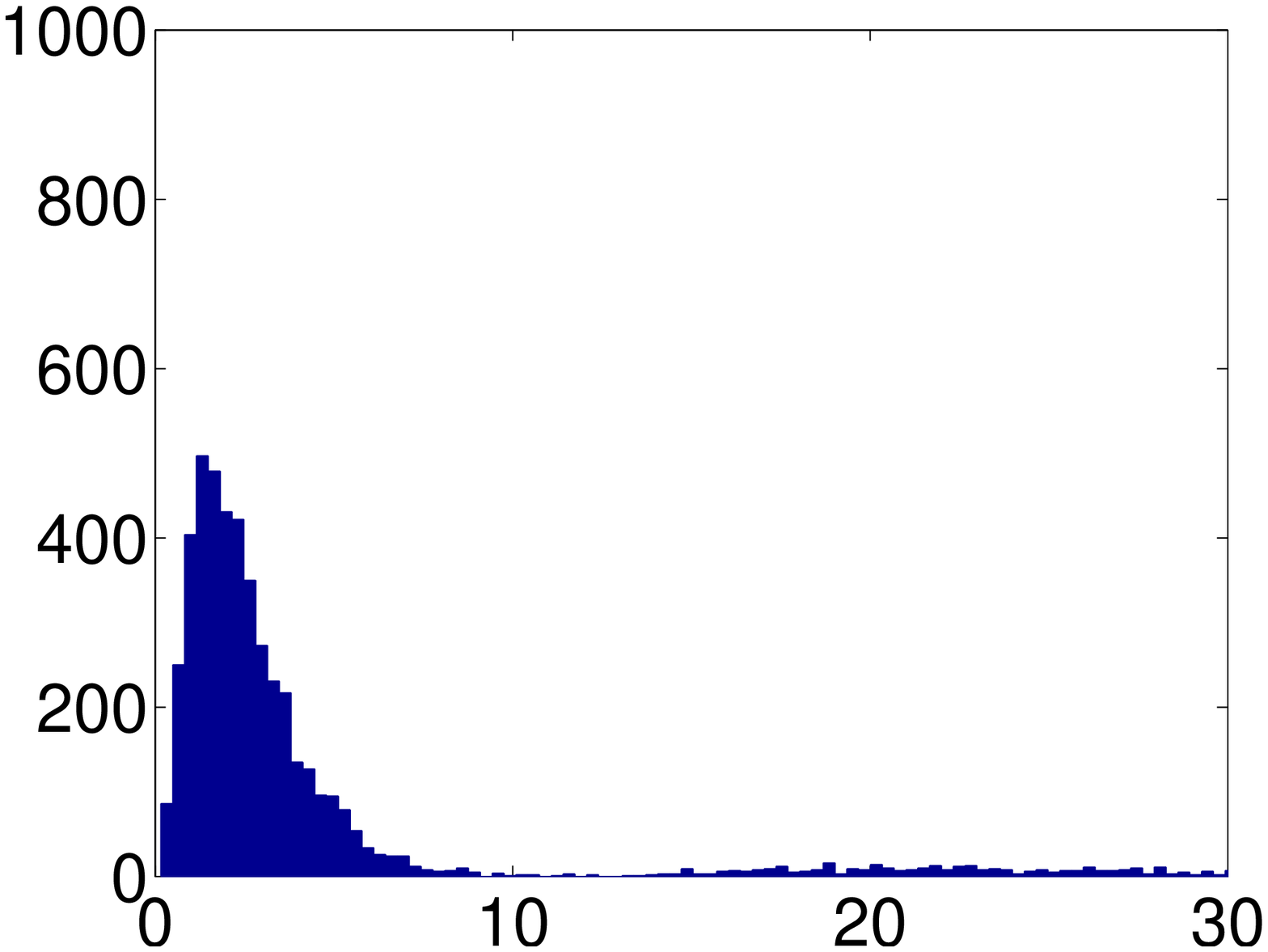} &
\includegraphics[scale=0.4]{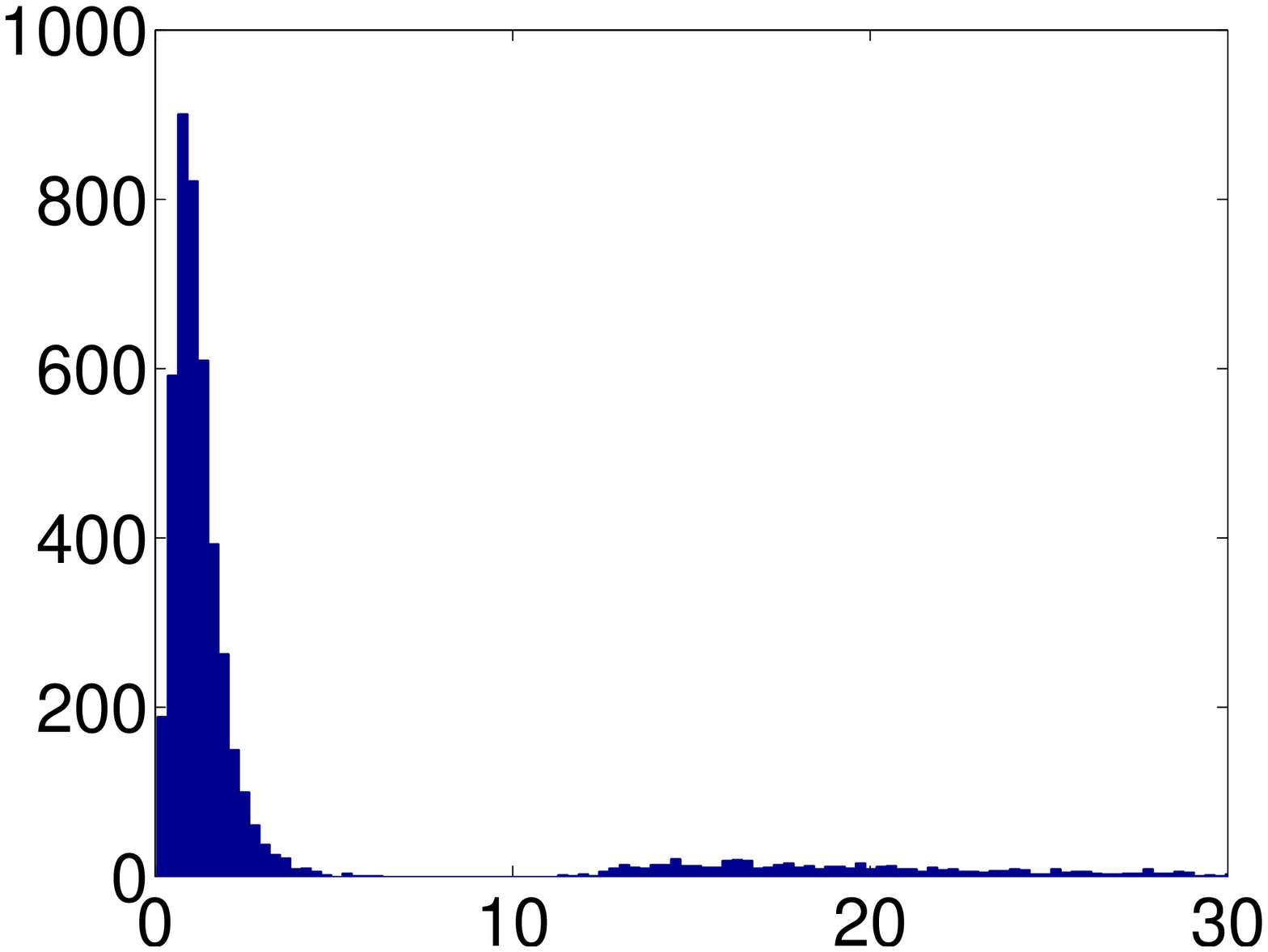} \\
(a) & (b)
\end{tabular}
\end{center} \caption{\small\sl Histogram of the ratio $\rho^2$
between the squared error loss and the ideal squared error for
(a)~unweighted Gauss-Dantzig estimator and (b)~reweighted
Gauss-Dantzig estimator. Approximately $5\%$ of the tail of each
histogram has been truncated for display; across 5000 trials the
maximum value observed was $\rho^2 \approx
165$.}\label{fig:rwdz2}
\end{figure}

We repeat the above experiment across 5000 trials.
Figure~\ref{fig:rwdz2} shows a histogram of the ratio $\rho^2$ between
the squared error loss of some estimate $x$ and the ideal squared
error
\[
\rho^2 := \frac{\sum_{i =1}^\ncol (x_i- x_{0,i})^2}{\sum_{i =
  1}^\ncol \min(x_{0,i}^2,\sigma^2)}
\]
for both the unweighted and reweighted Gauss-Dantzig estimators.
(The results are also summarized in Table~\ref{table:rwdz}.) For an
interpretation of the denominator, the ideal squared error $\sum
\min(x_{0,i}^2,\sigma^2)$ is roughly the mean-squared error one
could achieve if one had available an oracle supplying perfect
information about which coordinates of $\true$ are nonzero, and
which are actually worth estimating.  We see again a significant
reduction in reconstruction error; the median value of $\rho^2$
decreases from 2.43 to 1.21.  As pointed out, a primary reason for
this improvement comes from a more accurate identification of
significant coefficients: on average the unweighted Gauss-Dantzig
estimator includes 3.2 ``false positives,'' while the reweighted
Gauss-Dantzig estimator includes only 0.5. Both algorithms correctly
include all 8 nonzero positions in a large majority of trials.

\vfill

\begin{table}
\begin{center}
\begin{tabular}{| c | c | c |}
  \hline
  % after \\: \hline or \cline{col1-col2} \cline{col3-col4} ...
  ~ & Unweighted & Reweighted \\
  ~ & Gauss-Dantzig & Gauss-Dantzig \\
  \hline
  Median error ratio $\rho^2$ & 2.43  & 5.63 \\
  \hline
  Mean error ratio $\rho^2$ & 6.12 & 1.21 \\
  \hline
  Avg. false positives & 3.25 & 0.50 \\
  \hline
  Avg. correct detections & 7.86 & 7.80 \\
  \hline
\end{tabular}
\caption{Model selection results for unweighted and reweighted
versions of the Gauss-Dantzig estimator. In each of 5000 trials the
true sparse model contains $k=8$ nonzero entries.}
\label{table:rwdz}
\end{center}
\end{table}

\vfill

\subsection{Error correction}

Suppose we wish to transmit a real-valued signal $\true \in
\real^\ncol$, a block of $\ncol$ pieces of information, to a remote
receiver. The vector $\true$ is arbitrary and in particular,
nonsparse. The difficulty is that errors occur upon transmission so
that a fraction of the transmitted codeword may be corrupted in a
completely arbitrary and unknown fashion. In this setup, the authors
in \cite{CandesDLP} showed that one could transmit $n$ pieces of
information reliably by encoding the information as $\Phi \true$ where
$\Phi \in \R^{\nrow \times \ncol}$, $\nrow \ge \ncol$, is a suitable
coding matrix, and by solving
\begin{equation}
\label{eq:l1decode}
\min_{x \in \real^\ncol} \|y - \Phi x\|_{\ell_1}
\end{equation}
upon receiving the corrupted codeword $y = \Phi \true + e$; here, $e$
is the unknown but sparse corruption pattern. The conclusion of
\cite{CandesDLP} is then that the solution to this program recovers
$\true$ exactly provided that the fraction of errors is not too
large. Continuing on our theme, one can also enhance the performance
of this error-correction strategy, further increasing the number of
corrupted entries that can be overcome.

Select a vector of length $\ncol = 128$ with elements drawn from a
zero-mean unit-variance Gaussian distribution, and sample an $\nrow
\times \ncol$ coding matrix $\Phi$ with i.i.d.~Gaussian entries
yielding the codeword $\Phi x$.  For this experiment, $\nrow = 4\ncol
= 512$, and $\nsparse$ random entries of the codeword are corrupted
with a sign flip.  For the recovery, we simply use a reweighted
version of \eqref{eq:l1decode}. Our algorithm is as follows:
\begin{enumerate}
\item Set $\ell = 0$ and $w^{(0)}_i = 1$ for $i = 1, 2, \dots, \nrow$.
\item Solve the weighted $\ell_1$ minimization problem
\begin{equation}
\xhat^{(\ell)} = \arg\min \|W^{(\ell)} (y - \Phi \xhat)\|_{\ell_1}.
\label{eq:rwec}
\end{equation}
\item Update the weights; let $r^{(\ell)} = y - \Phi x^{(\ell)}$ and
  for each $i = 1, \ldots, \nrow$, define
\begin{equation}
  w^{(\ell+1)}_i = \frac{1}{|r^{(\ell)}_i|+\epsilon}.
\label{eq:rwruleEC}
\end{equation}
\item Terminate on convergence or when $\ell$ attains a specified
  maximum number of iterations $\ell_{\mathrm{max}}$. Otherwise,
  increment $\ell$ and go to step 2.
\end{enumerate}
We set $\epsilon$ to be some factor $\beta$ times the standard
deviation of the corrupted codeword $y$.  We run 100 trials for
several values of $\beta$ and of the size $\nsparse$ of the
corruption pattern.

\begin{figure}
\begin{center}
\includegraphics[scale=0.4]{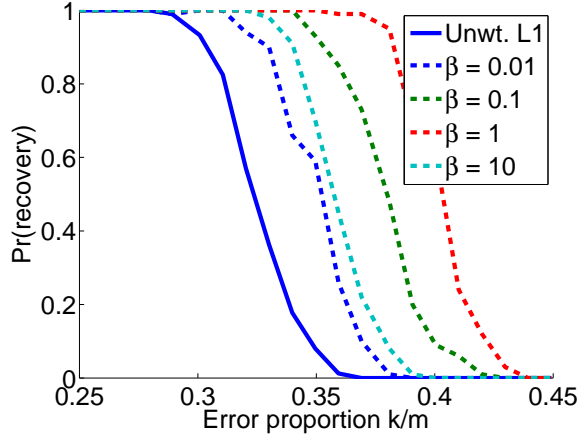}
\end{center}
\caption{\small\sl Unweighted (solid curve) and
  reweighted (dashed curve) $\ell_1$ signal recovery from corrupted
  measurements $y = \Phi \true + e$. The signal $\true$ has length $\ncol
  = 128$, the codeword $y$ has size $\nrow = 4\ncol = 512$, and the
  number of corrupted entries $\|e\|_{\ell_0} =
  \nsparse$.}\label{fig:ec}
\end{figure}

Figure~\ref{fig:ec} shows the probability of perfect signal recovery
(declared when $\|\true - \xhat\|_{\ell_\infty} \le 10^{-3}$) for
both the unweighted $\ell_1$ decoding algorithm and the reweighted
versions for various values of $\beta$ (with $\ell_{\mathrm{max}} =
4$). Across a wide range of values $\beta$ (and hence $\epsilon$),
we see that reweighting increases the number of corrupted entries
(as a percentage of the codeword size $\nrow$) that can be overcome,
from approximately $28\%$ to $35\%$.

\subsection{Total variation minimization for sparse image gradients}

In a different direction, reweighting can also boost the performance
of total-variation (TV) minimization for recovering images with sparse
gradients. Recall the total-variation norm of a 2-dimensional array
$(x_{i,j})$, $1 \le i, j \le n$, defined as the $\ell_1$ norm of the
magnitudes of the discrete gradient,
\begin{equation*}
\|x\|_{\textrm{TV}} = \sum_{1 \le i,j \le n-1} \|(D x)_{i,j}\|,
\end{equation*}
where $(D x)_{i,j}$ is the 2-dimensional vector of forward
differences $(Dx)_{i,j} = (x_{i+1,j} - x_{i,j}, x_{i,j+1} -
x_{i,j})$.  Because many natural images have a sparse or nearly
sparse gradient, it makes sense to search for the reconstruction
with minimal TV norm, i.e.,
\begin{equation}
  \min ~\|x\|_{\textrm{TV}} \quad \textrm{subject to} \quad y = \Phi x
\label{eq:tvopt};
\end{equation}
see~\cite{RudinOsherFatemi92,colorTV}, for example. It turns out
that this problem can be recast as a second-order cone
program~\cite{goldfarb05se}, and thus solved efficiently.

We adapt \eqref{eq:tvopt} by minimizing a sequence of weighted TV
norms as follows:
\begin{enumerate}
\item Set $\ell = 0$ and $w^{(0)}_{i,j} = 1$, $1 \le i,j \le \ncol-1$.
\item Solve the weighted TV minimization problem
\[
\xhat^{(\ell)} = \arg\min \, \sum_{1 \le i,j \le n-1} w^{(\ell)}_{i,j}
\, \|(D x)_{i,j}\|, \quad \textrm{subject to} \quad y = \Phi \xhat.
\]
\item Update the weights; for each $(i,j)$, $1 \le i, j \le \ncol -1$,
\begin{equation}
  w^{(\ell+1)}_{i,j} = \frac{1}{\|(D \xhat^{(\ell)})_{i,j}\|+\epsilon}.
\label{eq:rwruleTV}
\end{equation}
\item Terminate on convergence or when $\ell$ attains a specified
  maximum number of iterations $\ell_{\mathrm{max}}$. Otherwise,
  increment $\ell$ and go to step 2.
\end{enumerate}
Naturally, this iterative algorithm corresponds to minimizing a
sequence of linearizations of the log-sum function $\sum_{1 \le i,j
  \le n-1} \log(\|(D x)_{i,j}\|+\epsilon)$ around the previous signal
estimate.

\begin{figure}
\begin{center}
\begin{tabular}{cccc}
\includegraphics[scale=0.23]{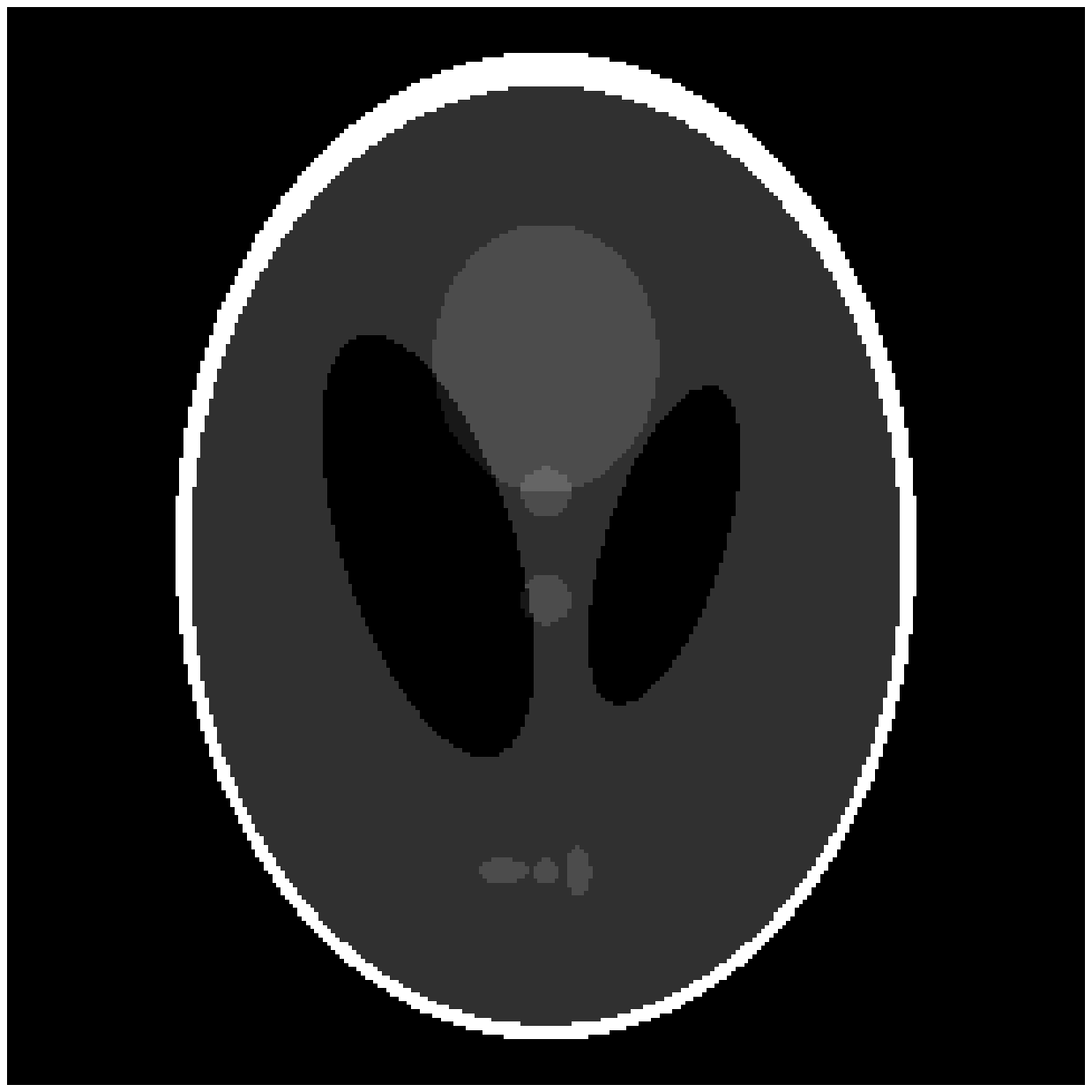} &
\includegraphics[scale=0.23]{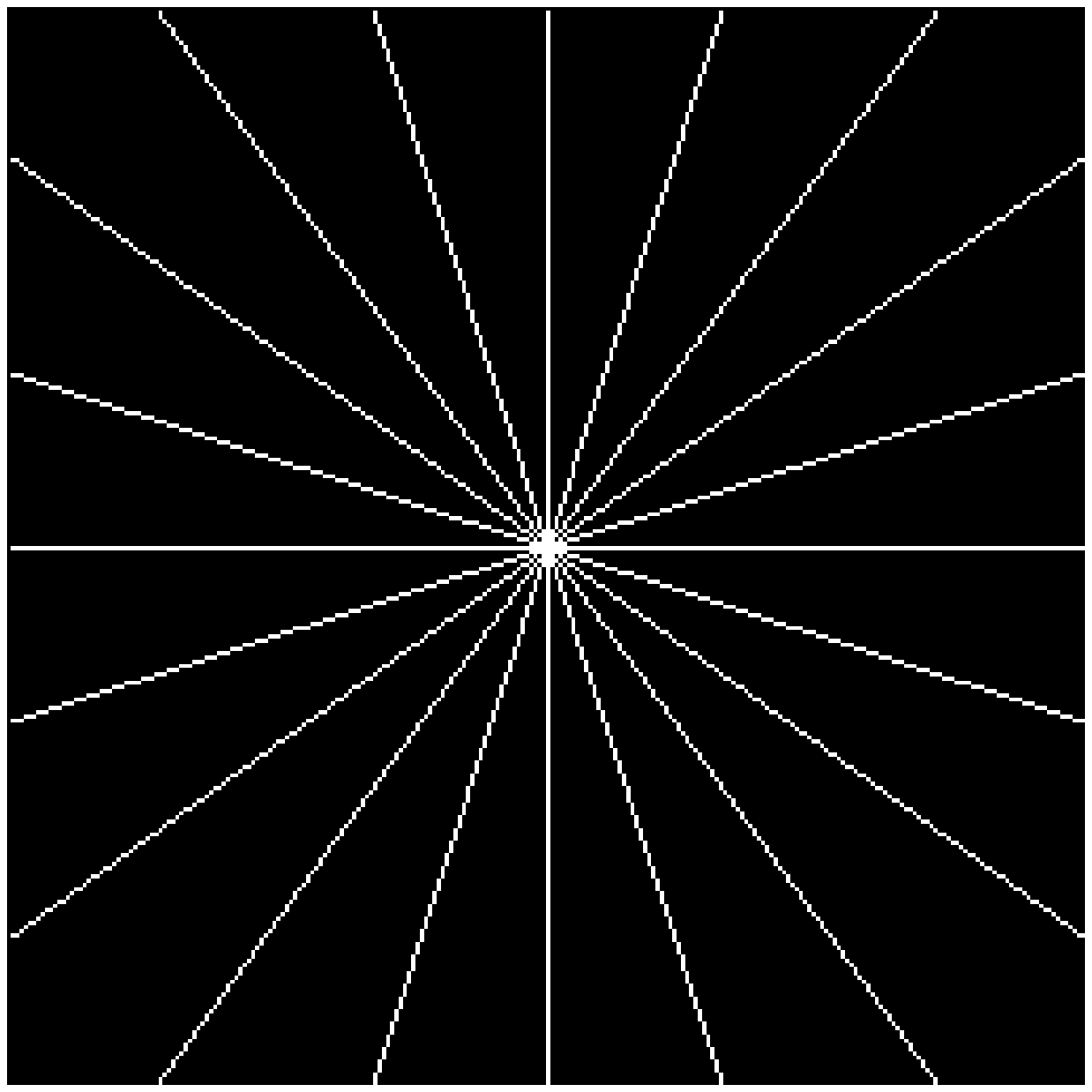} &
\includegraphics[scale=0.23]{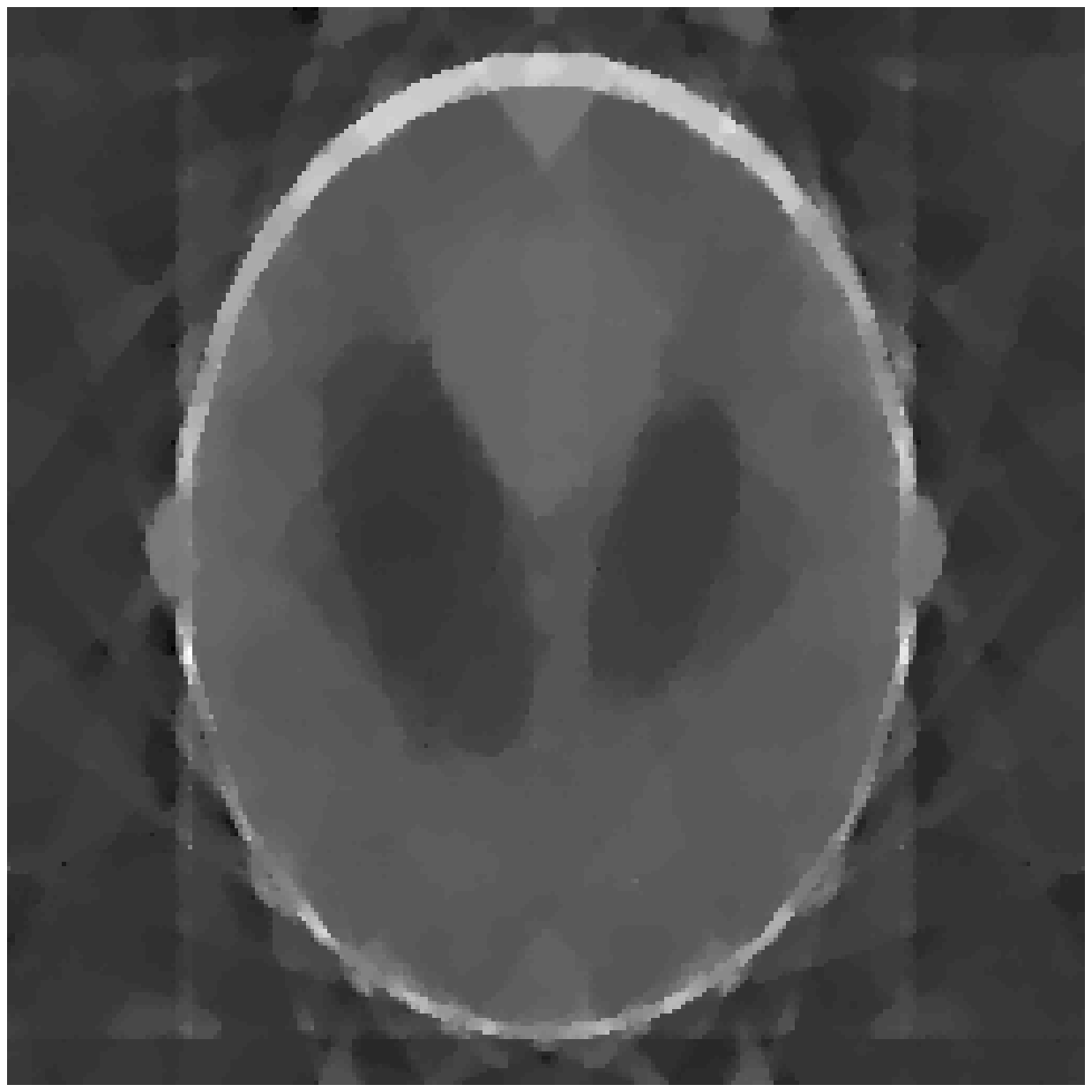} &
\includegraphics[scale=0.23]{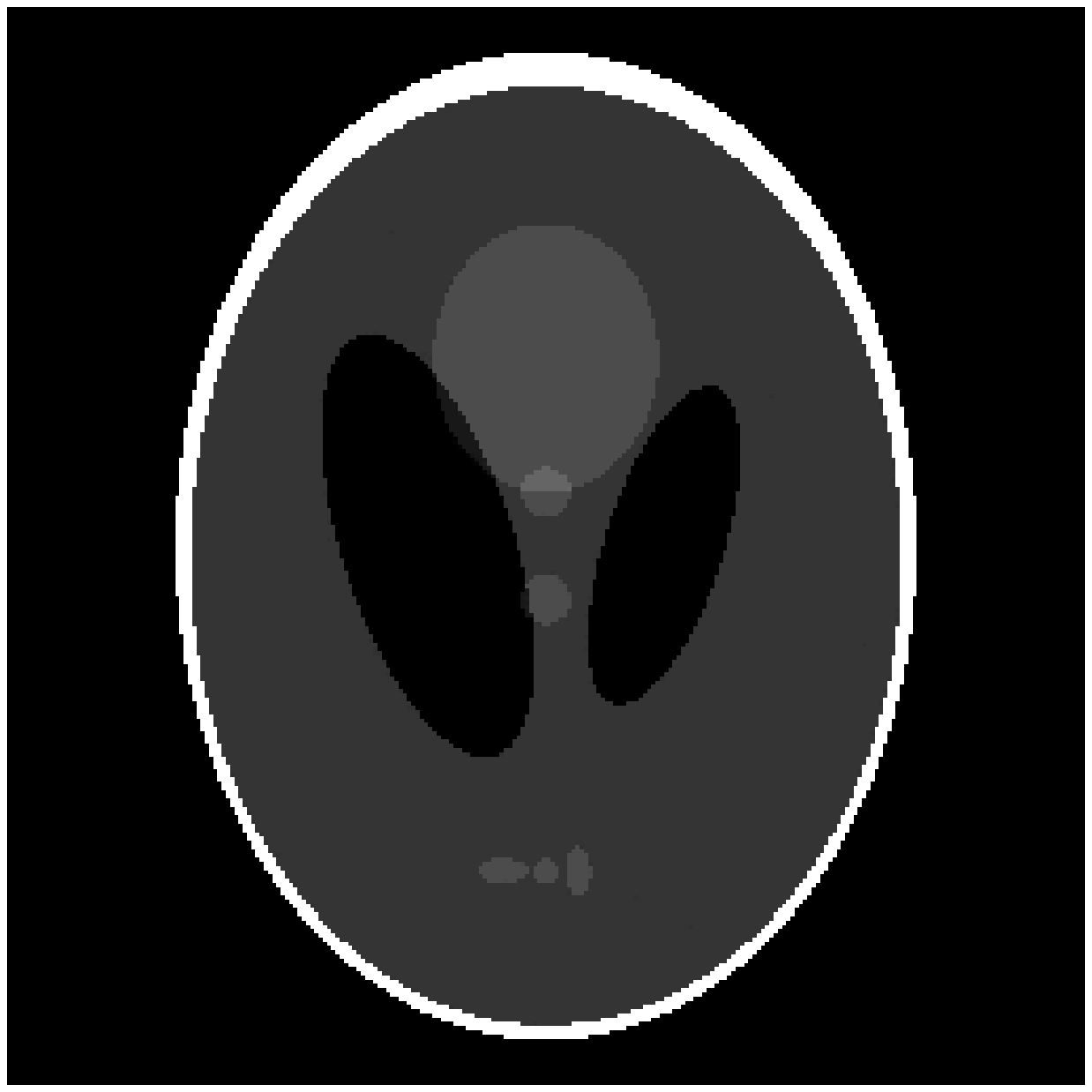} \\
(a) & (b) & (c) & (d)
\end{tabular}
\end{center}
\caption{\small\sl Image recovery from reweighted TV
minimization. (a) Original $256\times 256$ phantom image. (b)
Fourier-domain sampling pattern. (c) Minimum-TV reconstruction;
total variation = 1336. (d) Reweighted TV reconstruction; total
variation (unweighted) = 1464.}\label{fig:phantom}
\end{figure}

To show how this can enhance the performance of the recovery, consider
the following experiment. Our test image is the Shepp-Logan phantom of
size $\ncol = 256 \times 256$ (see Figure~\ref{fig:phantom}(a)). The
pixels take values between $0$ and $1$, and the image has a nonzero
gradient at 2184 pixels.  We measure $y$ by sampling the discrete
Fourier transform of the phantom along $10$ pseudo-radial lines (see
Figure~\ref{fig:phantom}(b)).  That is, $y = \Phi \true$, where $\Phi$
represents a subset of the Fourier coefficients of $\true$. In total,
we take $\nrow = 2521$ real-valued measurements.

Figure~\ref{fig:phantom}(c) shows the result of the classical
TV minimization, which gives a relative error equal to $\|\true -
x^{(0)}\|_{\ell_2}/\|\true\|_{\ell_2} \approx 0.43$. As shown in
Figure~\ref{fig:phantom}(d), however, we see a substantial improvement
after 6 iterations of reweighted TV minimization (we used $0.1$ for
the value of $\epsilon$). The recovery is near-perfect, with a
relative error obeying $\|\true -
x^{(6)}\|_{\ell_2}/\|\true\|_{\ell_2} \approx 2\times10^{-3}$.

For point of comparison it takes approximately $17$ radial lines
($\nrow = 4257$ real-valued measurements) to perfectly recover the
phantom using unweighted TV minimization. Hence, with respect to the
sparsity of the image gradient, we have reduced the requisite
oversampling factor significantly, from $\frac{4257}{2184} \approx
1.95$ down to $\frac{2521}{2184} \approx 1.15$. It is worth noting
that comparable reconstruction performance on the phantom image has
also been recently achieved by directly minimizing a nonconvex
$\ell_p$ norm, $p < 1$, of the image gradient~\cite{chartrand07ex};
we discuss this approach further in Section~\ref{sec:related}.

\section{Reweighted $\ell_1$ analysis}
\label{sec:l1analysis}

In many problems, a signal may assume sparsity in a possibly
overcomplete representation. To make things concrete, suppose we are
given a dictionary $\Psi$ of waveforms $(\psi_j)_{j \in J}$
(the columns of $\Psi$) which allows representing any signal as $x =
\Psi \alpha$. The representation $\alpha$ is deemed sparse when the
vector of coefficients $\alpha$ has comparably few significant terms.
In some applications, it may be natural to choose $\Psi$ as an
orthonormal basis but in others, a sparse representation of the signal
$x$ may only become possible when $\Psi$ is a {\em redundant}
dictionary; that is, it has more columns than rows. A good example is
provided by an audio signal which often is sparsely represented
as a superposition of waveforms of the general shape $\sigma^{-1/2}
g((t-t_0)/\sigma) e^{i \omega t}$, where $t_0$, $\omega$, and $\sigma$
are discrete shift, modulation and scale parameters.

In this setting, the common approach for sparsity-based recovery from
linear measurements goes by the name of {\em Basis Pursuit}
\cite{DonohoBP} and is of the form
\begin{equation}
  \min~ \|\alphahat\|_{\ell_1} \quad \textrm{subject to}
\quad  y = \Phi \Psi \alphahat;
\label{eq:ell1syn}
\end{equation}
that is, we seek a sparse set of coefficients $\alphahat$ that
synthesize the signal $\xhat = \Psi \alphahat$. We call this {\em
  synthesis-based $\ell_1$ recovery}. A far less common approach,
however, seeks a signal $\xhat$ whose coefficients $\alphahat =
\Psi^\ast \xhat$ (when $\xhat$ is analyzed in the dictionary $\Psi$)
are sparse
\begin{equation}
  \min \|\Psi^\ast \xhat\|_{\ell_1}
\quad \textrm{subject to} \quad y = \Phi \xhat.
  \label{eq:ell1anl}
\end{equation}
We call this {\em analysis-based $\ell_1$ recovery}. When $\Psi$ is an
orthonormal basis, these two programs are identical, but in general
they find {\em different} solutions. When $\Psi$ is redundant,
(\ref{eq:ell1anl}) involves fewer unknowns than (\ref{eq:ell1syn}) and
may be computationally simpler to solve~\cite{starck04re}.  Moreover,
in some cases the analysis-based reconstruction may in fact be
superior, a phenomenon which is not very well understood;
see~\cite{elad07an} for some insights.

Both programs are amenable to reweighting but what is interesting is
the combination of analysis-based $\ell_1$ recovery and iterative
reweighting which seems especially powerful. This section provides two
typical examples. For completeness, the iterative reweighted
$\ell_1$-analysis algorithm is as follows:
\begin{enumerate}
\item Set $\ell = 0$ and $w^{(\ell)}_j = 1$, $j \in J$ ($J$ indexes
  the dictionary).

\item Solve the weighted $\ell_1$ minimization problem
\[
\xhat^{(\ell)} = \arg\min \|W^{(\ell)} \Psi^\ast \xhat\|_{\ell_1} \quad
\textrm{subject to} \quad y = \Phi \xhat.
\]
\item Put $\alphahat^{(\ell)} = \Psi^\ast \xhat^{(\ell)}$ and
  define
\[
w_j^{(\ell+1)} = \frac{1}{|\alphahat^{(\ell)}_j|+\epsilon}, \quad j
\in J.
\]
\item Terminate on convergence or when $\ell$ attains a specified
  maximum number of iterations $\ell_{\mathrm{max}}$. Otherwise,
  increment $\ell$ and go to step 2.
\end{enumerate}

\subsection{Incoherent sampling of radar pulses}

Our first example is motivated by our own research focused on
advancing devices for analog-to-digital conversion of high-bandwidth
signals. To cut a long story short, standard analog-to-digital
converter (ADC) technology implements the usual quantized Shannon
representation; that is, the signal is uniformly sampled at or above
the Nyquist rate. The hardware brick wall is that conventional
analog-to-digital conversion technology is currently limited to sample
rates on the order of 1GHz, and hardware implementations of high
precision Shannon-based conversion at substantially higher rates seem
out of sight for decades to come. This is where the theory of
compressive sensing becomes relevant.

Whereas it may not be possible to digitize an analog signal at a very
high rate rate, it may be quite possible to change its polarity at a
high rate. The idea is then to multiply the signal by a pseudo-random
sequence of plus and minus ones, integrate the product over time
windows, and digitize the integral at the end of each time
interval. This is a parallel architecture and one has several of these
random multiplier-integrator pairs running in parallel using distinct
or event nearly independent pseudo-random sign sequences.

\begin{figure}
\begin{center}
\begin{tabular}{cc}
\includegraphics[scale=0.3]{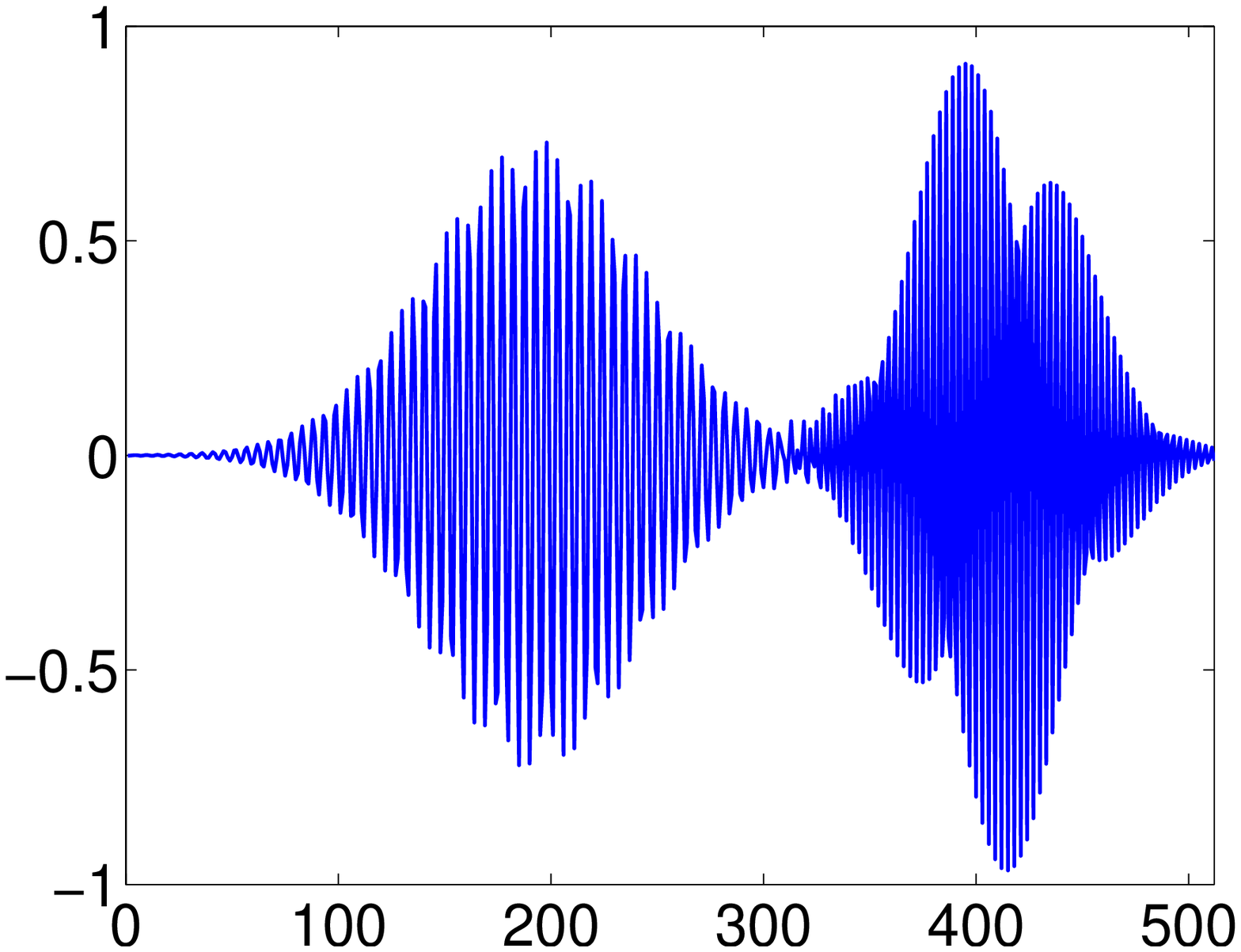} &
\includegraphics[scale=0.3]{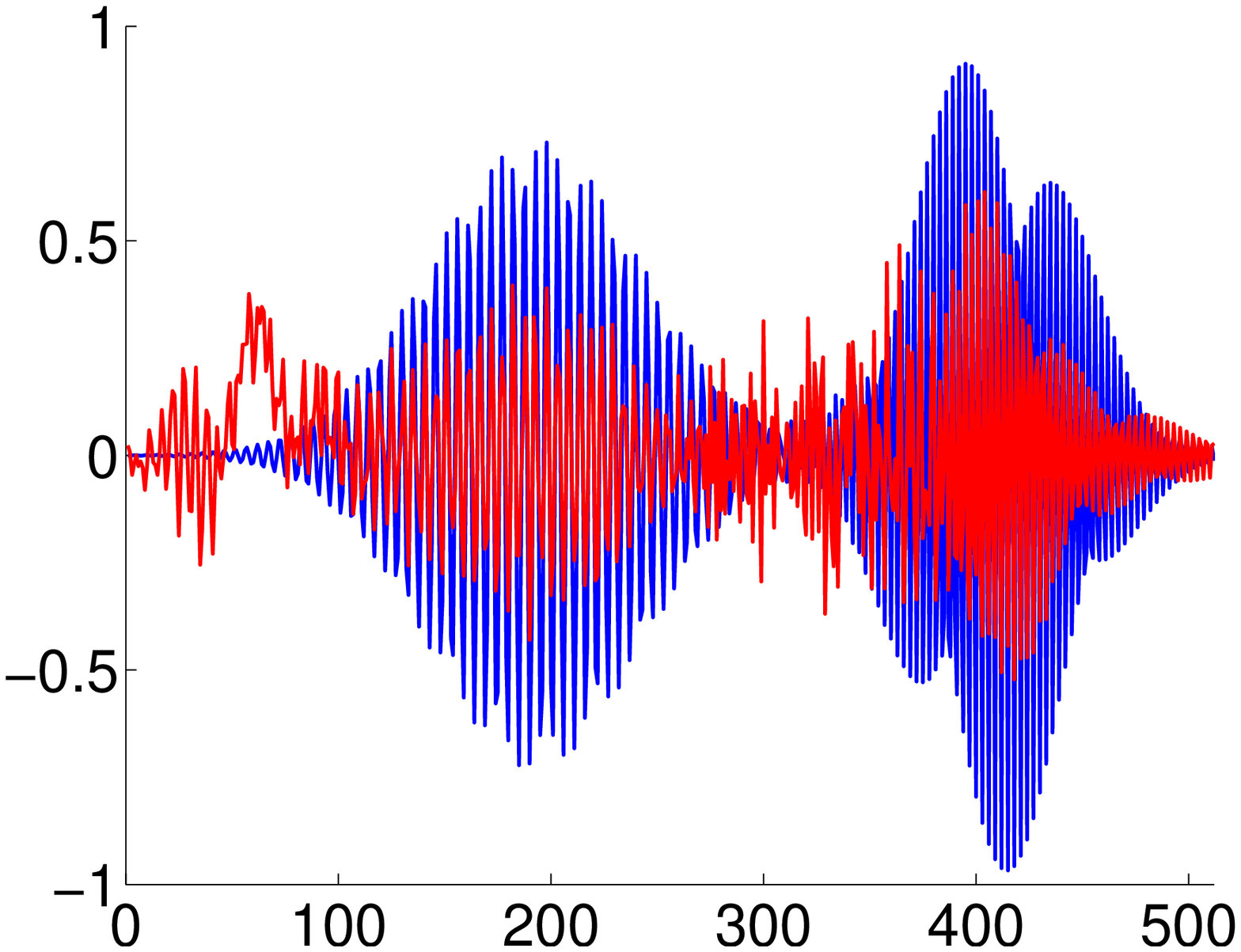}
\\
(a) & (b) \\
\includegraphics[scale=0.3]{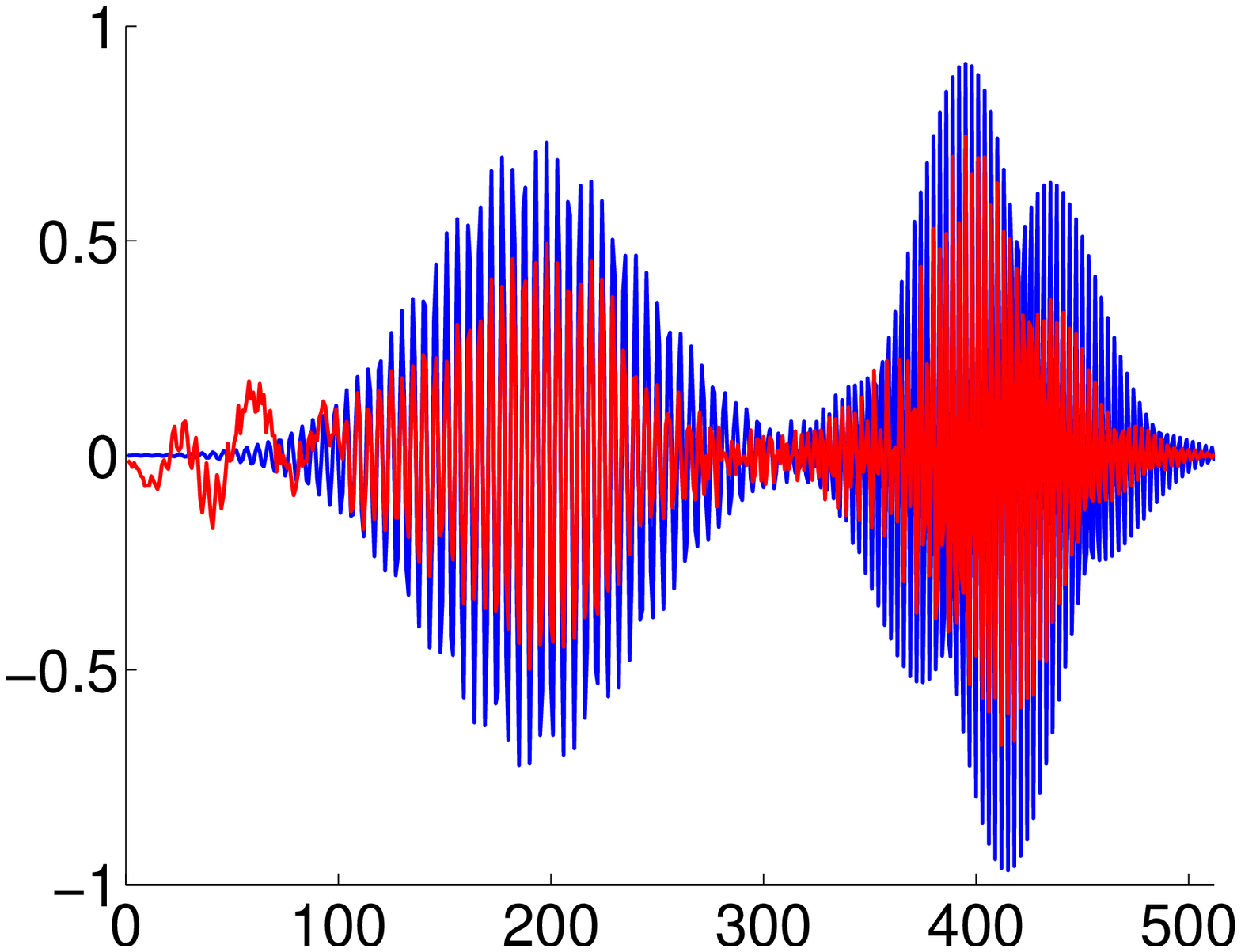} &
\includegraphics[scale=0.3]{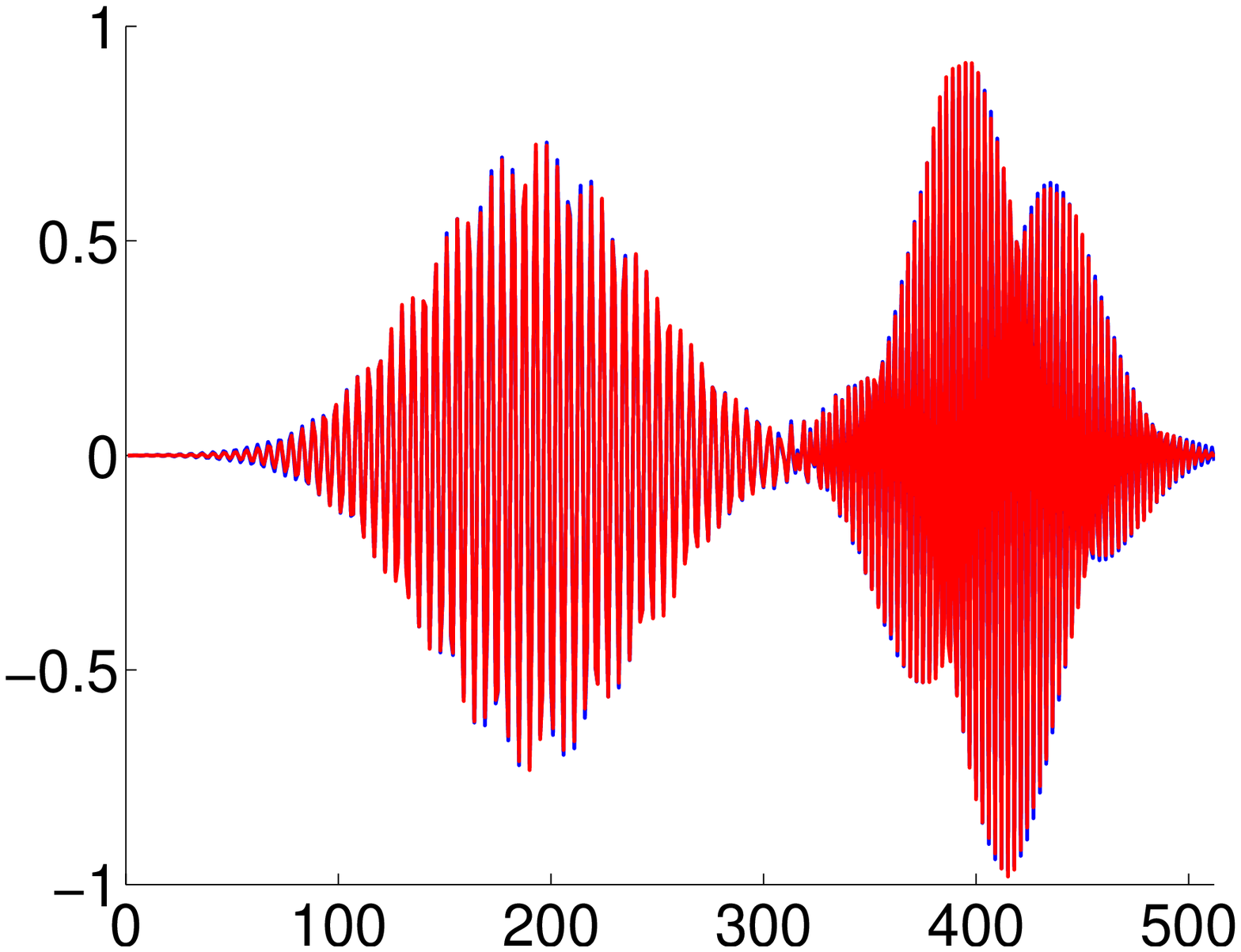} \\
(c) & (d)
\end{tabular}
\end{center}
\caption{\small\sl (a) Original two-pulse signal
(blue) and reconstructions (red) via (b) $\ell_1$ synthesis, (c)
$\ell_1$ analysis, (d) reweighted $\ell_1$ analysis. (e) Relative
$\ell_2$ reconstruction error as a function of reweighting
iteration.}\label{fig:pulse}
\end{figure}

To show the promise of this approach, we take $\true$ to be a 1-D
signal of length $\ncol=512$ which is a superposition of two
modulated pulses (see Figure~\ref{fig:pulse}(a)). From this signal,
we collect $\nrow = 30$ measurements using an $\nrow \times \ncol$
matrix $\Phi$ populated with i.i.d.~Bernoulli $\pm 1$ entries. This
is an unreasonably small amount of data corresponding to an
undersampling factor exceeding 17. For reconstruction we consider a
time-frequency Gabor dictionary that consists of a variety of sine
waves modulated by Gaussian windows, with different locations and
scales. Overall the dictionary is approximately $43\times$
overcomplete and does not contain the two pulses that comprise
$\true$.

Figure~\ref{fig:pulse}(b) shows the result of minimizing $\ell_1$
synthesis (\ref{eq:ell1syn}) in this redundant dictionary. The
reconstruction shows pronounced artifacts and
$\|\true-\xhat\|_{\ell_2}/\|x\|_{\ell_2} \approx 0.67$. These
artifacts are somewhat reduced by analysis-based $\ell_1$ recovery
(\ref{eq:ell1anl}), as demonstrated in Figure~\ref{fig:pulse}(c);
here, see $\|\true-\xhat\|_{\ell_2}/\|x\|_{\ell_2} \approx 0.46$.
However, reweighting the $\ell_1$ analysis problem offers a very
substantial improvement.  Figure~\ref{fig:pulse}(d) shows the result
after four iterations;
$\|\true-\xhat^{(4)}\|_{\ell_2}/\|x\|_{\ell_2}$ is now about 0.022.
Further, Table~\ref{table:pulse} shows the relative reconstruction
error $\|\true-\xhat^{(\ell)}\|_{\ell_2}/\|x\|_{\ell_2}$ as a
function of the iteration count $\ell$. Massive gains are achieved
after just 4 iterations.

\begin{table}
\begin{center}
\begin{tabular}{| c || c | c | c | c | c | c | c | c |}
  \hline
  Iteration count $\ell$ & 0 & 1 & 2 & 3 & 4 & 5 & 6 & 7 \\
  \hline
  Error  $\|\true-\xhat^{(\ell)}\|_{\ell_2}/\|x\|_{\ell_2}$ & 0.460 & 0.101 & 0.038 & 0.024 & 0.022 & 0.022 & 0.022 &
  0.022 \\
  \hline
\end{tabular}
\caption{Relative $\ell_2$ reconstruction error as a function of
reweighting iteration for two-pulse signal reconstruction.}
\label{table:pulse}
\end{center}
\end{table}

\subsection{Frequency sampling of biomedical images}

Compressed sensing can help reduce the scan time in Magnetic Resonance
Imaging (MRI) and offer sharper images of living tissues. This is
especially important because time consuming MRI scans have
traditionally limited the use of this sensing modality in important
applications. Simply put, faster imaging here means novel
applications. In MR, one collects information about an object by
measuring its Fourier coefficients and faster acquisition here means
fewer measurements.

\begin{figure}
\begin{center}
\begin{tabular}{cc}
\includegraphics[scale=0.45]{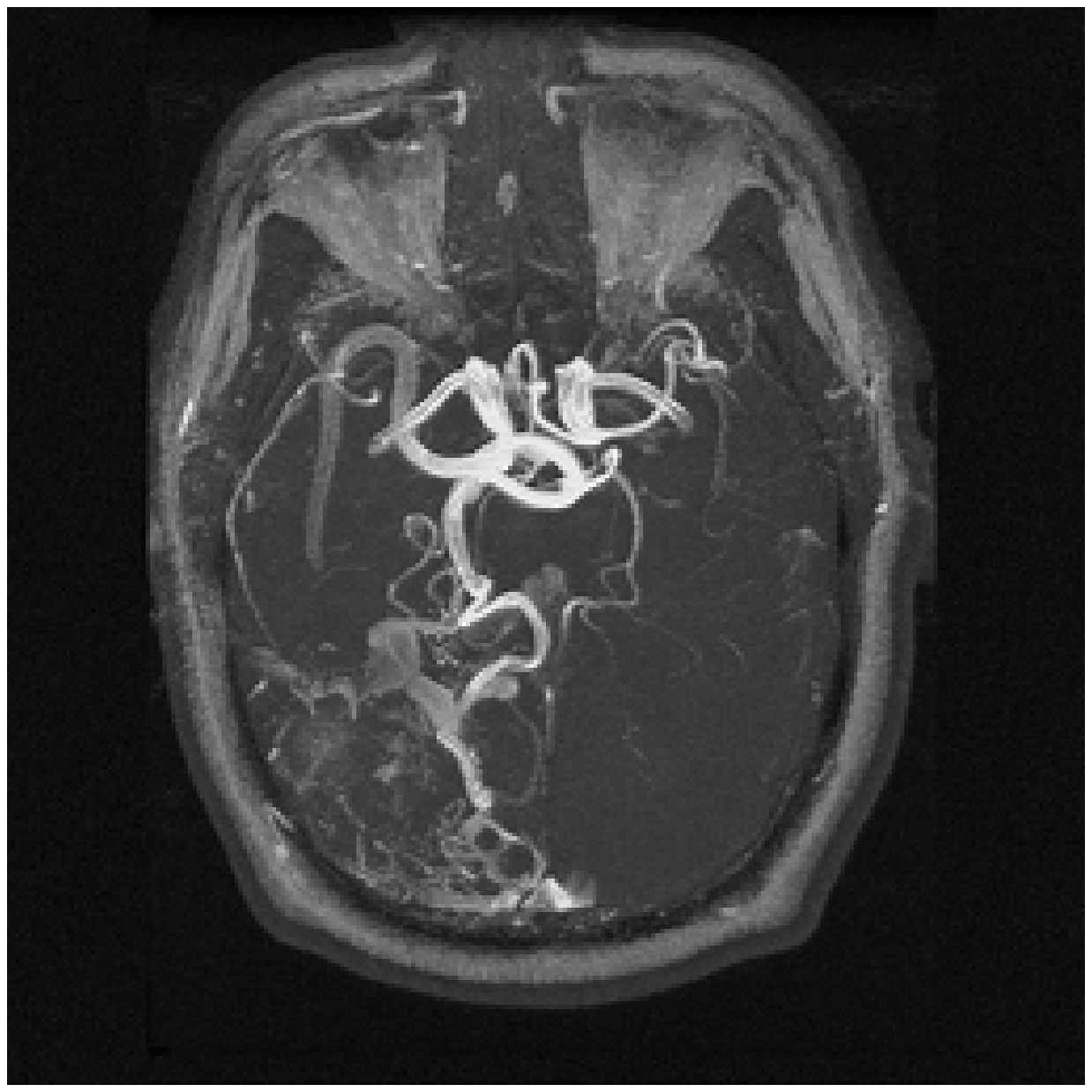} &
\includegraphics[scale=0.45]{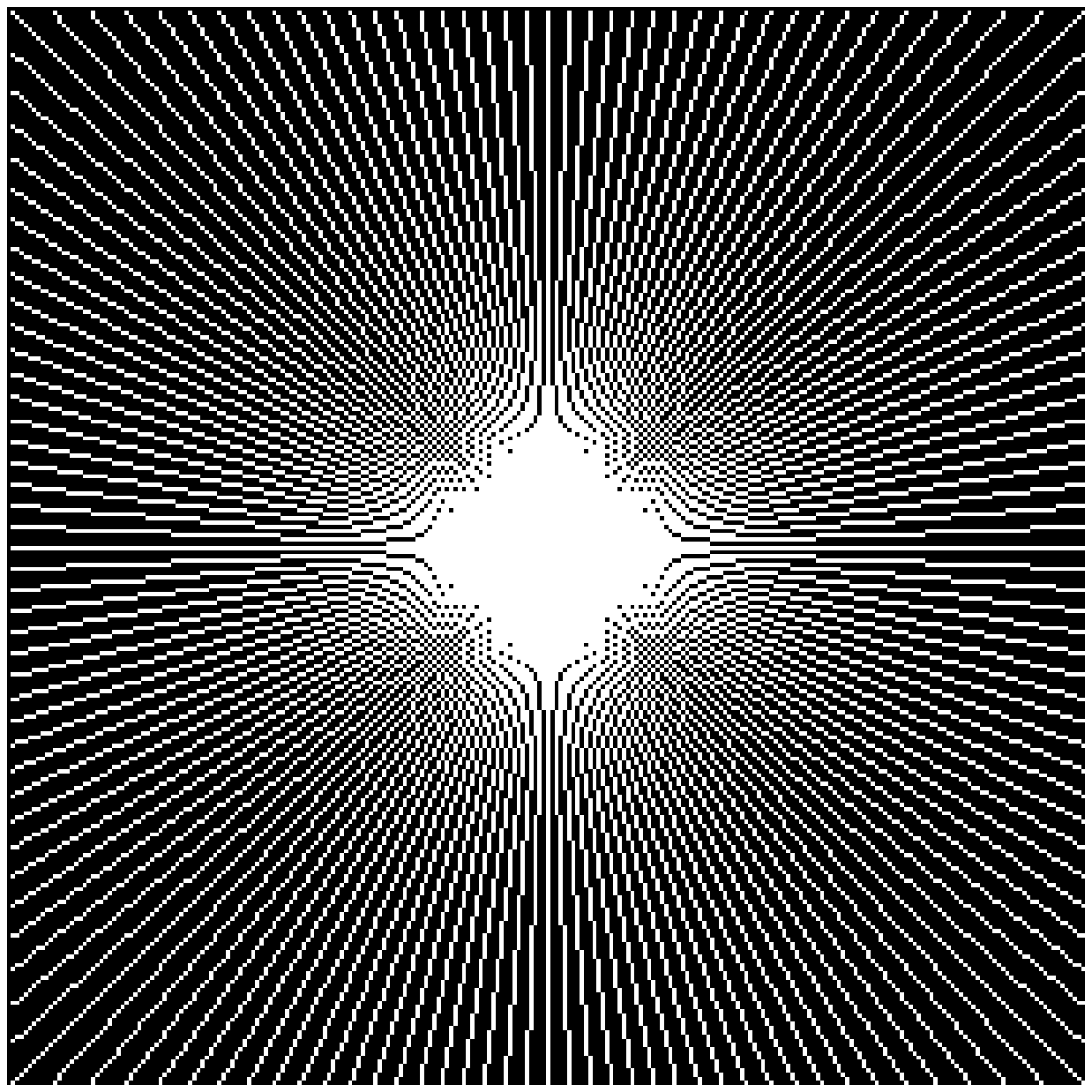} \\
(a) & (b)\\
\includegraphics[scale=0.45]{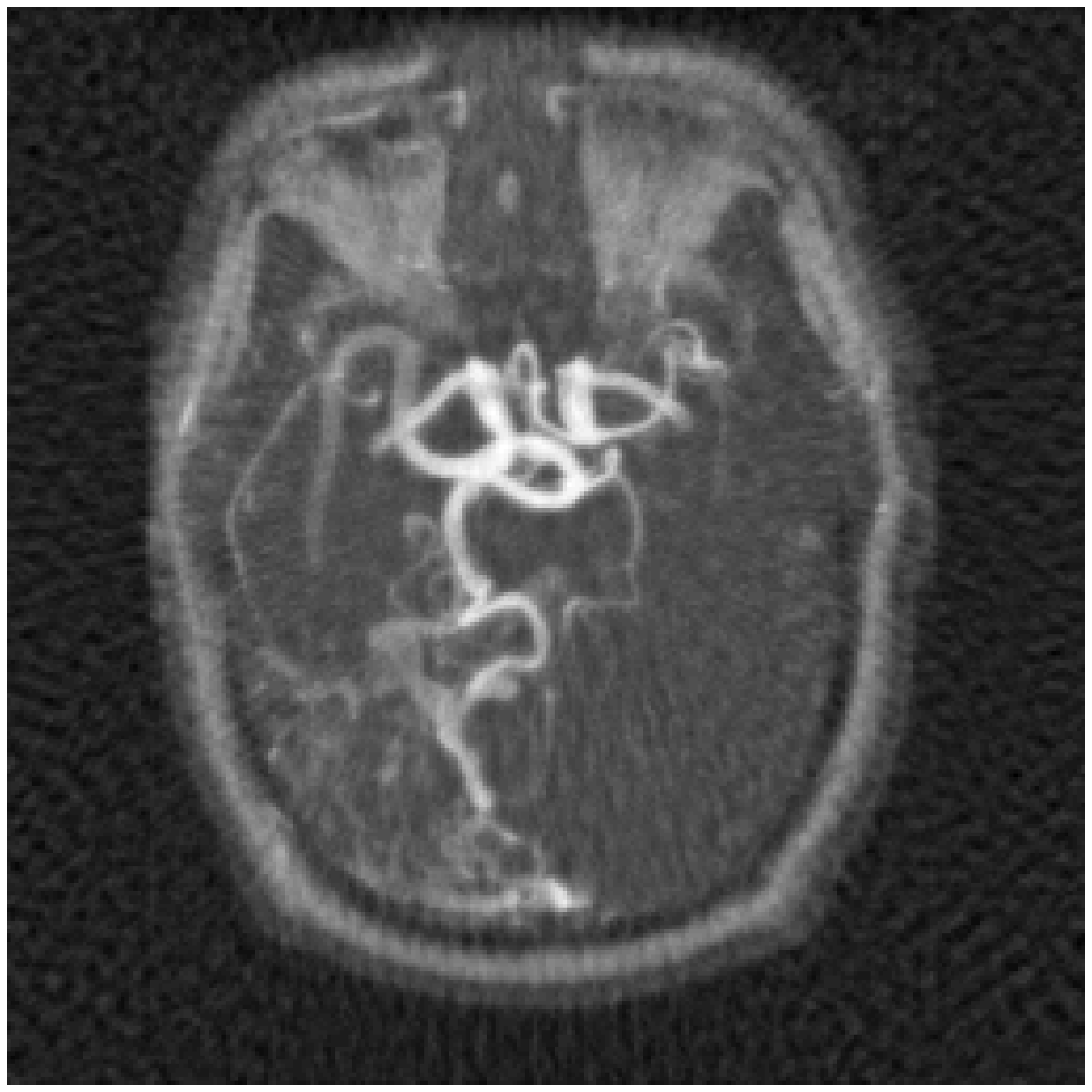} &
\includegraphics[scale=0.45]{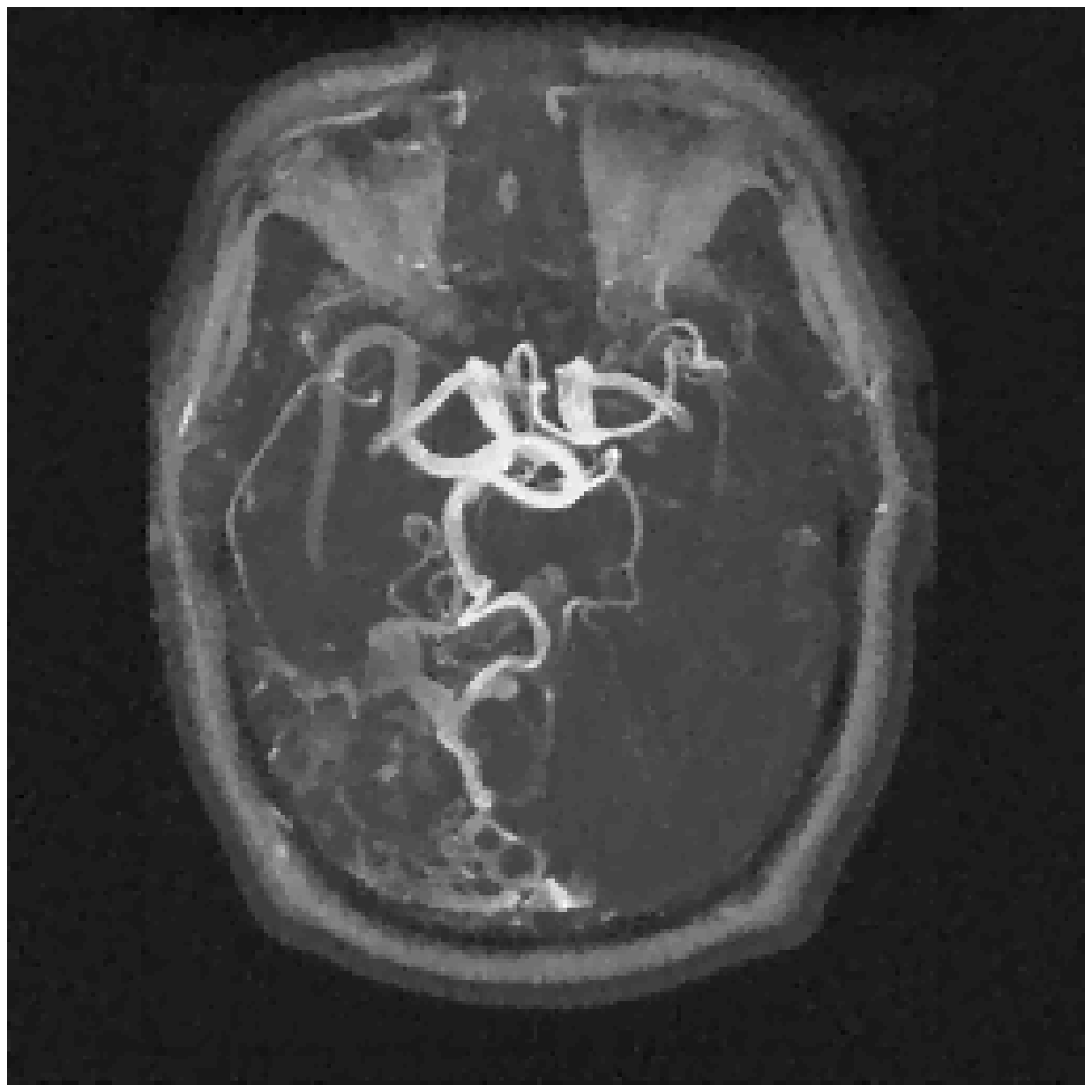} \\
(c) & (d) \\
\includegraphics[scale=0.45]{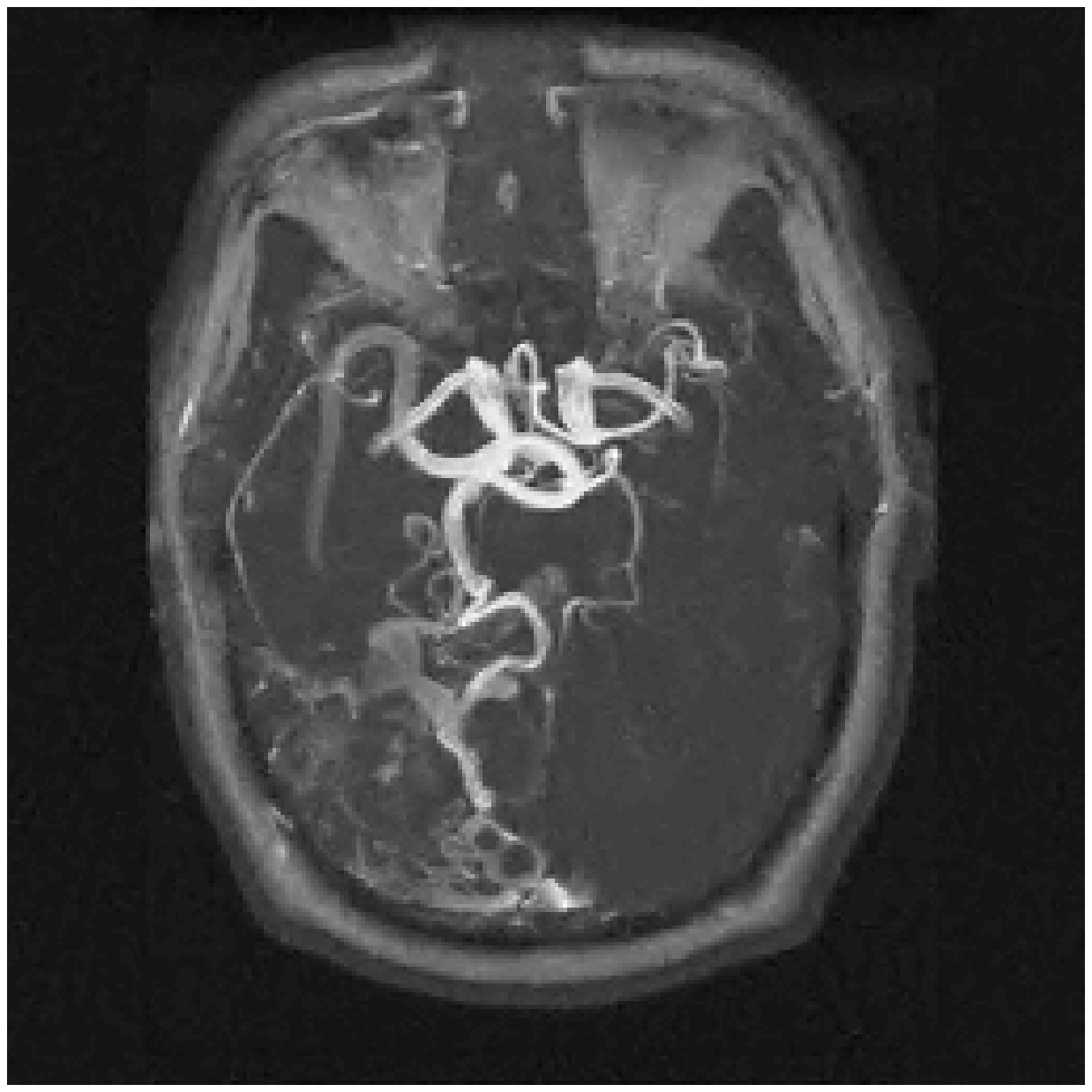} &
\includegraphics[scale=0.45]{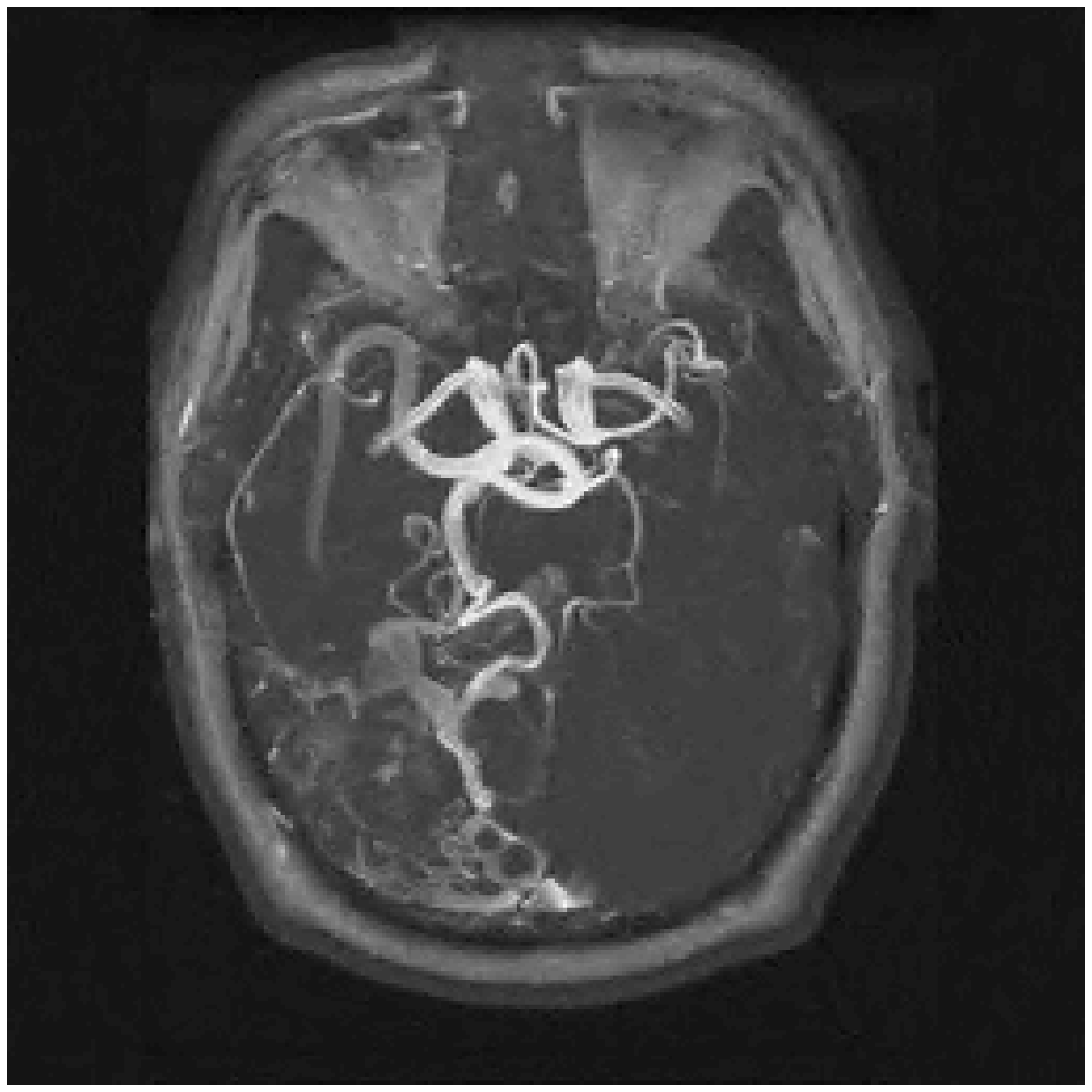} \\
(e) & (f)
\end{tabular}
\end{center}
\caption{\small\sl (a) Original MR angiogram. (b)
Fourier sampling pattern. (c) Backprojection, PSNR = 29.00dB. (d)
Minimum TV reconstruction, PSNR = 34.23dB. (e) $\ell_1$ analysis
reconstruction, PSNR = 34.37dB. (f)~Reweighted $\ell_1$ analysis
reconstruction, PSNR = 34.78dB.}\label{fig:angio}
\end{figure}

We mimic an MR experiment by taking our unknown image $\true$ to be
the $\ncol = 256 \times 256 = 65536$ pixel MR angiogram image shown
in Figure~\ref{fig:angio}(a).  We sample the image along 80 lines in
the Fourier domain (see Figure~\ref{fig:angio}(b)), effectively
taking $\nrow = 18737$ real-valued measurements $y = \Phi \true$. In
plain terms, we undersample by a factor of about 3.

Figure~\ref{fig:angio}(c) shows
the minimum energy reconstruction which solves
\begin{equation}
  \min \|\xhat\|_{\ell_2} \quad \textrm{subject to} \quad  y = \Phi \xhat.
\label{eq:ell2opt}
\end{equation}
Figure~\ref{fig:angio}(d) shows the result of TV minimization.  The
minimum $\ell_1$-analysis \eqref{eq:ell1anl} solution where $\Psi$
is a three-scale redundant D4 wavelet dictionary that is $10$ times
overcomplete, is shown on Figure~\ref{fig:angio}(e).
Figure~\ref{fig:angio}(f) shows the result of reweighting the
$\ell_1$ analysis with $\ell_{\mathrm{max}} = 4$ and $\epsilon$ set
to 100. For a point of comparison, the maximum wavelet coefficient
has amplitude 4020, and approximately 108000 coefficients (out of
655360) have amplitude greater than 100.

We can reinterpret these results by comparing the reconstruction
quality to the best $\nsparse$-term approximation to the image
$\true$ in a nonredundant wavelet dictionary. For example, an
$\ell_2$ reconstruction error equivalent to the $\ell_2$
reconstruction of Figure~\ref{fig:angio}(c) would require keeping
the $\nsparse = 1905 \approx \nrow/9.84$ largest wavelet
coefficients from the orthogonal wavelet transform of our test
image. In this sense, the requisite oversampling factor can be
thought of as being $9.84$. Of course this can be substantially
improved by encouraging sparsity, and the factor is reduced to
$3.33$ using TV minimization, to $3.25$ using $\ell_1$ analysis, and
to $3.01$ using reweighted $\ell_1$ analysis.

We would like to be clear about what this means. Consider the image in
Figure~\ref{fig:angio}(a) and its best $k$-term wavelet approximation
with $k = 6225$; that is, the approximation obtained by computing all
the D4 wavelet coefficients and retaining the $k$ largest in the
expansion of the object (and throwing out the others). Then we have
shown that the image obtained by measuring $3k$ real-valued Fourier
measurements and solving the iterative reweighted $\ell_1$ analysis
has just about the same accuracy. That is, the oversampling factor
needed to obtain an image of the same quality as if one knew ahead of
time the locations of the $k$ most significant pieces of information
and their value, is just 3.

\section{Discussion}
\label{sec:discussion}

In summary, reweighted $\ell_1$ minimization outperforms plain
$\ell_1$ minimization in a variety of setups. Therefore, this
technique might be of interest to researchers in the field of
compressed sensing and/or statistical estimation as it might help to
improve the quality of reconstructions and/or estimations.  Further,
this technique is easy to deploy as (1) it can be built on top of
existing $\ell_1$ solvers and (2) the number of iterations is
typically very low so that the additional computational cost is not
prohibitive. We conclude this paper by discussing related work and
possible future directions.

\subsection{Related work}
\label{sec:related}

Whereas we have focused on modifying the $\ell_1$ norm, a number of
algorithms been have proposed that involve successively reweighting
alternative penalty functions. In addition to IRLS (see
Section~\ref{sec:irls}), several such algorithms deserve mention.

Gorodnitsky and Rao~\cite{focuss} propose FOCUSS as an iterative
method for finding sparse solutions to underdetermined systems. At
each iteration, FOCUSS solves a reweighted $\ell_2$ minimization
with weights
\begin{equation}
w^{(\ell)}_i = \frac{1}{{\xhat^{(\ell-1)}_i}} \label{eq:wfocuss}
\end{equation}
for $i = 1,2,\dots,\ncol$. For nonzero signal coefficients, it is
shown that each step of FOCUSS is equivalent to a step of the
modified Newton's method for minimizing the function
\begin{equation*}
\sum_{i : \xhat_i \neq 0} \log |\xhat_i|
\end{equation*}
subject to $y = \Phi \xhat$. As the iterations proceed, it is
suggested to identify those coefficients apparently converging to
zero, remove them from subsequent iterations, and constrain them
instead to be identically zero.

In a small series of experiments, we have observed that reweighted
$\ell_1$ minimization recovers sparse signals with lower error (or
from fewer measurements) than the FOCUSS algorithm. We attribute
this fact, for one, to the natural tendency of unweighted $\ell_1$
minimization to encourage sparsity (while unweighted $\ell_2$
minimization does not). We have also experimented with an
$\epsilon$-regularization to the reweighting function
(\ref{eq:wfocuss}) that is analogous to (\ref{eq:rwrule}). However
we have found that this formulation fails to encourage strictly
sparse solutions. (Sparse solutions can be encouraged by letting
$\epsilon \rightarrow 0$ as the iterations proceed, but the overall
performance remains inferior to reweighted $\ell_1$ minimization
with fixed $\epsilon$.)

Harikumar and Bresler~\cite{harikumar96ne} propose an iterative
algorithm that can be viewed as a generalization of FOCUSS. At each
stage, the algorithm solves a convex optimization problem with a
reweighted $\ell_2$ cost function that encourages sparse solutions.
The algorithm allows for different reweighting rules; for a given
choice of reweighting rule, the algorithm converges to a local minimum
of some concave objective function (analogous to the log-sum penalty
function in~(\ref{eq:logsum})). These methods build upon $\ell_2$
minimization rather than $\ell_1$ minimization.

Delaney and Bresler~\cite{delaney98gl} also propose a general
algorithm for minimizing functionals having concave regularization
penalties, again by solving a sequence of reweighted convex
optimization problems (though not necessarily $\ell_2$ problems) with
weights that decrease as a function of the prior estimate. With the
particular choice of a log-sum regularization penalty, the algorithm
resembles the noise-aware reweighted $\ell_1$ minimization discussed
in Section~\ref{sec:qcl1}.

Finally, in a slightly different vein, Chartrand~\cite{chartrand07ex}
has recently proposed an iterative algorithm to minimize the concave
objective $\|\xhat\|_{\ell_p}$ with $p<1$. (The algorithm alternates
between gradient descent and projection onto the constraint set $y =
\Phi \xhat$.) While a global optimum cannot be guaranteed, experiments
suggests that a local minimum may be found---when initializing with
the minimum $\ell_2$ solution---that is often quite sparse. This
algorithm seems to outperform $(\Pone)$ in a number of instances and
offers further support for the utility of nonconvex penalties in
sparse signal recovery. To reiterate, a major advantage of reweighted
$\ell_1$ minimization in this thrust is that (1) it can be implemented
in a variety of settings (see Sections~\ref{sec:exp}
and~\ref{sec:l1analysis}) on top of existing and mature linear
programming solvers and (2) it typically converges in very few
steps. The log-sum penalty is also more $\ell_0$-like and as we
discuss in Section~\ref{sec:variations}, additional concave penalty
functions can be considered simply by adapting the reweighting rule.

\subsection{Future directions}

In light of the promise of reweighted $\ell_1$ minimization, it seems
desirable to further investigate the properties of this algorithm.
\begin{itemize}
\item Under what conditions does the algorithm converge? That is, when
  do the successive iterates $\xhat^{(\ell)}$ have a limit
  $\xhat^{(\infty)}$?

\item As shown in Section \ref{sec:overview}, when there is a sparse
  solution and the reweighted algorithm finds it, convergence may
  occur in just very few steps. It would be of interest to understand
  this phenomenon more precisely.

\item What are smart and robust rules for selecting the parameter
  $\epsilon$? That is, rules that would automatically adapt to the
  dynamic range and the sparsity of the object under study as to
  ensure reliable performance across a broad array of signals. Of
  interest are ways of updating $\epsilon$ as the algorithm progresses
  towards a solution. Of course, $\epsilon$ does not need to be
  uniform across all coordinates.

\item We mentioned the use of other functionals and reweighting rules.
  How do they compare?

\item Finally, any result quantifying the improvement of the
  reweighted algorithm for special classes of sparse or nearly sparse
  signals would be significant.
\end{itemize}

\subsection*{Acknowledgments}
E.~C. was partially supported by a National Science Foundation grant
CCF-515362, by the 2006 Waterman Award (NSF) and by a grant from
DARPA. This work was performed while M.~W. was an NSF Postdoctoral
Fellow (NSF DMS-0603606) in the Department of Applied and
Computational Mathematics at Caltech. S.~B. was partially supported
by NSF award 0529426, NASA award NNX07AEIIA, and AFOSR awards
FA9550-06-1-0514 and FA9550-06-1-0312. We would like to thank
Nathaniel Braun and Peter Stobbe for fruitful discussions about this
project. Parts of this work were presented at the Fourth IEEE
International Symposium on Biomedical Imaging (ISBI `07) held April
12--15, 2007 and at the Von Neumann Symposium on Sparse
Representation and High-Dimensional Geometry held July 8--12, 2007.
Related work was first developed as lecture notes for the course
\emph{EE364b: Convex Optimization II}, given at Stanford Winter
quarter 2006-07 \cite{EE364b:07}.

\clearpage
{\small
\bibliographystyle{IEEEbib}
\bibliography{IEEEabrv,rwl1bibOct2007}
}
\end{document}